\documentclass[a4paper,11pt]{article}
\pdfoutput=1

\usepackage{jheppub}
\usepackage[utf8]{inputenc}
\usepackage{amsmath, amssymb}
\usepackage{mathrsfs}
\usepackage{float}
\usepackage{color}
\usepackage{xcolor}
\usepackage{multirow}
\usepackage{graphicx,caption,subcaption}
\usepackage{bm}
\usepackage{dsfont}
\usepackage{cancel}
\usepackage{latexsym}
\usepackage{slashed}
\usepackage[space]{grffile}
\usepackage{tikz}
\usepackage{enumitem}

\newcommand{\Ga}{{\Gamma}}
\newcommand{\De}{{\Delta}}

\newcommand{\al}{{\alpha}}
\newcommand{\bt}{{\beta}}
\newcommand{\ga}{{\gamma}}
\newcommand{\de}{{\delta}}
\newcommand{\ep}{{\epsilon}}
\newcommand{\vep}{{\varepsilon}}
\newcommand{\zt}{{\zeta}}
\newcommand{\te}{{\theta}}

\newcommand{\vphi}{{\varphi}}

\newcommand{\mrm}[1]{{\mathrm{#1}}}

\newcommand{\bsk}{{\boldsymbol{k}}}

\newcommand{\bsn}{{\boldsymbol{n}}}

\newcommand{\bsx}{{\boldsymbol{x}}}
\newcommand{\bsy}{{\boldsymbol{y}}}

\newcommand{\bsnl}{{\boldsymbol{0}}}

\newcommand{\bbR}{{\mathbb{R}}}

\newcommand{\bbZ}{{\mathbb{Z}}}

\newcommand{\cA}{{\mathcal{A}}}
\newcommand{\cB}{{\mathcal{B}}}
\newcommand{\cC}{{\mathcal{C}}}
\newcommand{\cD}{{\mathcal{D}}}

\newcommand{\cH}{{\mathcal{H}}}
\newcommand{\cI}{{\mathcal{I}}}

\newcommand{\cK}{{\mathcal{K}}}
\newcommand{\cL}{{\mathcal{L}}}

\newcommand{\cO}{{\mathcal{O}}}
\newcommand{\cP}{{\mathcal{P}}}

\newcommand{\cV}{{\mathcal{V}}}

\newcommand{\rd}{{\mathrm{d}}}

\newcommand{\mcd}{{\kern-0.1pt\cdot\kern-0.1pt}}
\newcommand{\nab}{{\nabla}}
\newcommand{\nn}{{\nonumber}}

\newcommand{\ol}{\overline}
\newcommand{\pd}{{\partial}}

\newcommand{\bra}{{\langle}}
\newcommand{\ket}{{\rangle}}

\newcommand{\bbra}{{\langle\kern-2.5pt\langle}}
\newcommand{\kket}{{\rangle\kern-2.5pt\rangle}}
\newcommand{\Bbra}{{\Big\langle\kern-3.5pt\Big\langle}}
\newcommand{\Kket}{{\Big\rangle\kern-3.5pt\Big\rangle}}

\DeclareMathOperator*{\ssum}{\mbox{$\sum$}}
\DeclareMathOperator*{\sprod}{\mbox{$\prod$}}

\title{Bispectrum from five-dimensional inflation}

\author[a,b]{Ignatios Antoniadis}
\author[a]{Auttakit Chatrabhuti}
\author[b]{Jules Cunat}
\author[a]{Hiroshi Isono}

\affiliation[a]{High Energy Physics Research Unit, Faculty of Science,
Chulalongkorn University,\\
Bangkok 10330, Thailand}
\affiliation[b]{Laboratoire de Physique Th\'eorique et Hautes Energies - LPTHE\\
Sorbonne Universit\'e, CNRS, 4 Place Jussieu, 75005 Paris, France}

\emailAdd{antoniad@lpthe.jussieu.fr, auttakit.c@chula.ac.th, jcunat@lpthe.jussieu.fr, hiroshi.i@chula.ac.th}

\abstract{It was proposed that five-dimensional (5D) inflation can blow up the size of a compact dimension from the 5D Planck length to the micron size, as required by the dark dimension proposal, relating the weakness of the actual gravitational force to the size of the observable universe. Moreover, it was shown that 5D inflation can generate the (approximate) flat power spectrum of primordial density fluctuations consistent with present observations. Here we compute the bispectrum of primordial scalar perturbations and show that unlike the power spectrum, it differs from the four-dimensional case at all angular distances, due to the fact that in contrast to global dilatations, invariance under special conformal transformations is not restored at late times.
Moreover there is an additional enhancement in the squeezed limit.}

\begin{document}
\maketitle



\section{Introduction}\label{sec:Intro}
Extra dimensions is a general prediction of string theory and there are good physics motivations that their size may be larger than the Planck length, accounting for large scale hierarchies in Nature~\cite{Antoniadis:1990ew, Arkani-Hamed:1998jmv, Antoniadis:1998ig}. A natural question is then why and how extra dimensions can obtain large size. A possible answer that was recently proposed is via higher dimensional homogeneous inflation that can blow them up from a fundamental length to the one required for the observed strength of gravity, together with expanding the size of the observable universe as required for explaining the horizon problem~\cite{Anchordoqui:2022svl, Anchordoqui:2023etp}. Higher dimensional inflation can then relate two large hierarchies in particle physics and cosmology, namely the weakness of the gravitational force and the largeness of the observable universe, starting from the  scale of a microscopic theory, such as string theory. 

It turns out that higher dimensional inflation also leads to (approximate up to slow-roll parameters) scale invariant spectrum of primordial density perturbations on a 3-brane universe at distances less than the compactification length at the end of inflation, while it deviates for larger distances~\cite{Anchordoqui:2023etp}. Consistency with CMB observations then implies that the size of extra dimensions should be
larger than the scales probed accurately by cosmological observations that correspond to angular distances of about $0.01 - 10$ degrees in the sky, associated to physical distances of order kpc to Mpc at the time of CMB. Extrapolating these scales back in the early universe at the end of inflation using radiation dominated expansion and translating them to higher dimensional Planck units, one finds that the Mpc becomes about a micron, which is a remarkable coincidence. As a result, if the size of extra dimensions is in a narrow window around the micron region (taking into account an upper bound of about 30 microns from the validity of Newton's law~\cite{Lee:2020zjt}), higher dimensional inflation can generate primordial perturbations consistent with observations. 
It follows that this mechanism can be nicely applied to the Dark Dimension proposal~\cite{Montero:2022prj} following the swampland distance conjecture~\cite{Ooguri:2006in} for the cosmological constant~\cite{Lust:2019zwm}, which implies an extra dimension at the micron scale associated to a five-dimensional (5D) Planck mass where gravity becomes strong of order $10^{9}$ GeV.\footnote{It was shown recently~\cite{Anchordoqui:2025nmb} that the case of two extra dimensions at the micron scale with a 6D Planck mass of order 10 TeV is also a viable (although marginally) possibility, surviving astrophysical and cosmological constraints due to the violation of momentum conservation in the compact directions~\cite{Hannestad:2003yd}.}

The power spectra of single field 5D inflation on a 3-brane universe originate from primordial perturbations of a scalar and the 5D graviton which can be decomposed from a four-dimensional (4D) point of view in terms of a spin-2 (B-modes), a spin-1 graviphoton and an additional scalar (the radion)~\cite{Antoniadis:2023sya}. The spin-2 is accompanied by a Kaluza-Klein (KK) tower of massive spin-2 states, while in $S^1/\mathbb{Z}_2$ brane world, the graviphoton has no 4D zero-mode. Similarly to 4D inflation, the graviton fluctuations are subdominant compared to the scalar by a slow-roll parameter, suppressing the effects of the second scalar and spin-1 polarisation and avoiding in particular possible iso-curvature perturbations unlike multi-field inflation. To allow comparison with 4D inflation, the slow-roll parameters are defined via time derivatives of the (slowing varying) Hubble parameter (Hubble-flow parameters) instead of field derivatives of the scalar potential.

Due to the localisation of a brane observer in the extra dimension, for a given three-dimensional momentum of the 3D non-compact dimensions one has to sum over all KK modes of the compact dimension, leading to a dependence of the power spectrum on the size of the extra dimension, or equivalently on the compactification scale. As a result, the observed power spectrum comes not only from a single field, the 4D zero-mode of the inflaton,  but  also from the whole tower of its massive KK excitations. It then turns out that (approximate up to slow-roll parameters) scale invariance is recovered in the limit of large 3D momenta, larger than the compactification scale, corresponding to angular distances less than about 10 degrees in the sky, consistently with CMB observations which are well measured in this region. On the other hand, at larger angular distances (small momenta), there is an important deviation from scale invariance leading to an enhancement of the power spectrum with vanishing spectral index, which could be tested in future observations along with other precision tests in cosmology.

In this work, we extend the above analysis by computing the bispectrum in single field 5D inflation, associated to  non-gaussianities, at lowest non-trivial order in the slow-roll parameters. We show that scale invariance is again restored when all distances are smaller than the compactification length corresponding to 10 degrees in the sky, while they deviate otherwise, like in the case of the power spectrum. As a result, in the momentum representation, the function $f_\text{NL}$ parametrising the non-gaussiantities depends on two ratios of 3D momenta, which however differs from the known result of the 4D single field inflation~\cite{Maldacena:2002vr}. This is the consequence of global special conformal symmetry breaking in inflation, unlike the global scale invariance restoration. A measurement of shape-dependent non-gaussiantites could therefore test 5D (or in general higher-dimensional) versus 4D single field inflation. Moreover, in the  squeezed limit of 3D momenta involving one distance larger than 10 degrees, there is an enhancement by 2-3 orders of magnitude compared to 4D inflation, due to the violation of global scale invariance.

The outline of the paper is the following. In Section~2, we present an overview of the framework of single field inflation in arbitrary dimensions. In Section~3, we describe the formalism for the computation of the cosmological correlators and review the results for the power spectrum in the bulk and on a 3-brane for convenience of the reader. In Section~4, we compute the bispectrum and its measurement by an observer on a 3-brane. In Section~5, we give the results for the non-gaussianity function $f_\text{NL}$ together with our concluding remarks. Finally, the paper contains two appendices. In Appendix~A, we describe the ADM formalism used for the computation and we present the full cubic action at all orders in Hubble-flow parameters, while in Appendix~B we study spatial global conformal invariance in 4D inflation.

\section{Single-field inflation in arbitrary dimensions}

In this section, we present an overview of single-field slow-roll inflation in arbitrary dimensions~\cite{Antoniadis:2023sya, Hirose:2025pzm}. We suppose that spacetime has $(1+d)$ dimensions and among $d$ spatial dimensions, three dimensions are non-compact $\bbR^3$ while each of the rest $(d-3)$ is compactified on a circle $S_1$, while our universe with the Standard Model is confined on a three-dimensional brane, located at a point in the compact space which we take for simplicity to be $(S_1)^{d-3}$. 

In the $(d+1)$-dimensional inflation period, the spacetime background is a $(d+1)$-dimensional quasi de Sitter (dS) spacetime in which the three non-compact and the $(d-3)$ compact dimensions expand exponentially with a common scale factor $\hat a$. Its metric is given by
\begin{align}\label{metric-bg}
\rd\bar s^2=\bar g_{\mu\nu}\rd x^\mu\rd x^\nu=-d\hat t^2+\hat a(\hat t)^2\rd\hat{\bsx}^2,
\end{align}
where $\hat\bsx=(\bsx,\bsx_\bot)$ denotes the comoving spatial coordinates, where $\bsx\in\bbR^3$ and $\bsx_\bot\in(S_1)^{d-3}$. We suppose that each direction $x_\bot^m$ ($m=4,\cdots,d$) of $\bsx_\bot$ ranges over $[0,2\pi R_0)$, describing a circle of radius $R_0$. We also suppose that the brane is located at $\bsx_\bot=\bsnl$.\footnote{In this case, the compactification space is in general an orbifold, such as a line segment for one dimension $S_1/\bbZ_2$, which brings only minor changes to our analysis.}

The full metric around the background \eqref{metric-bg} can be parametrised in the ADM form as
\begin{align}
\rd s^2=-N^2\rd\hat t^2+\hat a(\hat t)^2e^{2\zt}(e^\ga)_{ij}(\rd\hat x^i+N^i\rd\hat t)(\rd\hat x^j+N^j\rd\hat t),
\end{align}
where the indices $i,j$ range over $1,\cdots,d$. The functions $N,N^i$ are called the lapse and shift vector, respectively.
In this paper, we only consider the scalar fluctuation in $(d+1)$ dimensions $\zt$, thereby $\ga_{ij}=0$.\footnote{From the 4D point of view, the tensor fluctuation $\gamma_{ij}$ contains additional scalar modes besides the gravity waves that we neglect here. A dedicated analysis is needed to decide if their contribution is suppressed as in the power spectrum~\cite{Antoniadis:2023sya} or not.}
The conformal time $\tau$ is defined through $\rd\hat t=\hat a\rd\tau$.

We consider a model with a single scalar field (inflaton) $\phi$ in a potential $V(\phi)$ minimally coupled with the $(d+1)$-dimensional Einstein gravity. The action is given by
\begin{align}
S=\int\rd\hat t\rd^d\hat\bsx\sqrt{-g}\left[
\frac{M_*^{d-1}}{2}R(g)-\frac{1}{2}g^{\mu\nu}\pd_\mu\phi\pd_\nu\phi-V(\phi)
\right] \,.
\end{align}
We write the full configuration of the inflaton $\phi$ as the sum of the background trajectory $\phi_0(\hat t)$ and its fluctuation $\vphi$:
\begin{align}
\phi(\hat t,\hat\bsx)=\phi_0(\hat t)+\vphi(\hat t,\hat\bsx) \,.
\end{align}
The time-dependent background configurations $\hat a$ and $\phi_0$ are determined by the equations of motion (EoM):
\begin{align}
&d(d-1)M_*^{d-1}H^2=\dot\phi_0^2+2V, \\
&\dot\phi_0^2=-(d-1)M_*^{d-1}\dot H, \\
&\ddot\phi_0+dH\dot\phi_0+V'(\phi_0)=0, \quad 
\end{align}
where $V'=dV/d\phi$ and each dot is the $\hat t$-derivative. $H$ is the Hubble parameter defined by
\begin{align}
H(\hat t):=\frac{1}{\hat a}\frac{\rd\hat a}{\rd\hat t} \,.
\end{align}
Inflation occurs around a plateau of the scalar potential where spacetime is approximate de Sitter (dS) in Poincar\'e patch with flat compact directions. The deviation of this quasi-dS background from exact dS is characterised by small Hubble flow functions, slowly varied in time, defined by time derivatives of the Hubble parameter~\cite{Schwarz:2001vv}: 
\begin{align}
\vep_1:=-\frac{\dot H}{H^2}, \qquad \vep_{n+1}:=\frac{\dot\vep_n}{H\vep_n} \quad (n\geq 1) \,.
\end{align}
In terms of them, the scale factor $\hat a$ is given by $\hat a=-1/(H\tau)+\cO(\vep)$. 

\subsection{Action of scalar fluctuations}
Let us proceed to the full action in terms of the fluctuations. Since we will use it to compute expectation values of quantum fields, we have to fix the gauge of diffeomorphisms. As we focus on the scalar fluctuations $\zt,\vphi$, only one gauge degree of freedom is relevant and its fixing singles out only one physical scalar fluctuation. There are two popular gauge fixing conditions: one is the comoving gauge $\vphi=0$, and the other is the spatially flat gauge $\zt=0$.

In the case of our model, the lapse and shift vector $N,N^i$ satisfy algebraic equations of motion. Therefore, once the gauge is fixed, they can be solved order by order in the physical scalar fluctuation. Concretely, parametrising $N$ and $N_i=\hat a^2N^i$ as
\begin{align}
N=1+\al, \qquad N_i=\hat\pd_i\bt+\bt_i, \qquad \hat\pd_ib_i=0\,,
\end{align}
where $\hat\pd_i:=\pd/\pd\hat x^i$,
we solve the equations of motion \eqref{eq:constraint1} and \eqref{eq:constraint2} order by order in the fluctuation.
In order to obtain a cubic action in the fluctuation, it is sufficient to consider $\al,\bt$ and $b_i$ at the first order in the fluctuation. 

In this paper, we adopt $\zt$ as the dynamical variable, which is a natural choice since $\zt$ stays time-independent beyond the horizon scale. Since we will compute equal-time two- and three-point correlation functions of $\zt$, we need a cubic action in $\zt$. There are two ways to obtain it:
\begin{enumerate}
\item We choose the comoving gauge. The solutions $\al,\bt,b_i$ are given by ($\hat\pd^2=\hat\pd_i\hat\pd_i$)
\begin{align}
\al=\frac{1}{H}\dot\zt, \qquad  b_i=0, \qquad \hat\pd^2\bt=-\frac{1}{H}\hat\pd^2\zt+\frac{1}{d-1}\frac{a^2\dot\phi_0^2}{M_*^{d-1}H^2}\dot\zt.
\end{align}
While this is straightforward from the viewpoint of the dynamical variable, it involves huge number of integrations by part in order to reach a form of the action in which factors of slow-roll parameters are manifest. A full derivation of the cubic action in this gauge is given in~\cite{Collins:2011mz} for $d=3$.

\item We choose the spatially flat gauge. The solutions $\al,\bt,b_i$ are given by 
\begin{align}\label{NNisol-sfgauge}
\al=\frac{1}{d-1}\frac{\dot\phi_0}{M_*^{d-1}H}\vphi, \qquad  b_i=0, \qquad \hat\pd^2\bt=-\frac{1}{d-1}\frac{a^2\dot\phi_0^2}{M_*^{d-1}H^2}\frac{\rd}{\rd\hat t}\left(\frac{H}{\dot\phi_0}\vphi\right).
\end{align}
In this gauge, factors of slow-roll parameters are manifest from the beginning and hence complicated integrations by part are not required. We then change to the comoving gauge.
\end{enumerate}
In the following, we take the second path. First, we substitute the solutions \eqref{NNisol-sfgauge} into the action written in terms of the ADM variables \eqref{S-ADM}. We then obtain a cubic action $S_{\mrm{sf}}[\vphi]$ in which the coefficient of each cubic vertex is written in terms of first and second times derivatives of $\phi_0$ and derivatives of the potential $V$ and hence factors of the Hubble flow parameters can be made manifest by using the background equations of motion. 

The next step is to find  a coordinate transformation $\hat x_\vphi \mapsto \hat x_\zt$ that converts the gauges from the spatially flat $\zt=0$ to the comoving condition $\vphi=0$. Since the action is cubic, the transformation is at most of second order in fluctuations\footnote{The transformation at second order is needed since otherwise the condition of no metric fluctuation $\ga_{ij}=0$ cannot be kept.}. The procedure to obtain $\hat x_\vphi \mapsto \hat x_\zt$ and the consequent relation between $\vphi$ and $\zt$ is parallel to the $d=3$ case in~\cite{Maldacena:2002vr} and is summarised elsewhere such as in~\cite{Collins:2011mz,Antoniadis:2024abm}. We therefore just present the result. $\vphi$ in the spatially flat gauge and $\zt$ in the comoving gauge are related by
\begin{align}
\vphi=-\frac{\dot\phi_0}{H}\zt+\frac{\dot\phi_0}{H}f(\zt),
\end{align}
where $f(\zt)$ is given by
\begin{align}
f(\zt)=\frac{\vep_2}{4}\zt^2
+\frac{1}{H}\zt\dot\zt
+\frac{1}{2(d-1)\hat a^2H^2}
(\de_{ij}-\hat\pd^{-2}\hat\pd_i\hat\pd_j)(2\hat a^2H\hat\pd^{-2}\hat\pd_i\dot\zt\hat\pd_j\zt-\hat\pd_i\zt\hat\pd_j\zt)\,.
\label{f(zeta)}
\end{align}
Substituting this relation into the cubic action $S_{\mrm{sf}}[\vphi]$ in the spatially flat gauge, we obtain the cubic action in the comoving gauge:
\begin{align}
S_{\mrm{co}}[\zt]=\int\rd\hat t\rd^d\hat\bsx\,(\cL_2+\cL_3),
\end{align}
where the quadratic part $\cL_2$ is given by
\begin{align}
\cL_2&=(d-1)M_*^{d-1}\hat a^d\vep_1\left(\frac{1}{2}\dot\zt^2-\frac{1}{2}\hat a^{-2}\hat\pd_i\zt\hat\pd_i\zt\right) 
\end{align}
and the cubic interaction part $\cL_3$ is given by
\begin{align}
\cL_3&=\cV_0+\cV_1+\cV_2+\cV_3+\cO(\vep^3)
\end{align}
with the cubic vertices at the second order in the Hubble flow parameters given by
\begin{align}
\begin{split}
\cV_0&=M_*^{d-1}(d-1)[\pd_t(\hat a^d\vep_1\dot\zt)-\hat a^{d-2}\vep_1\hat\pd^2\zt]f(\zt), \\
\cV_1&=M_*^{d-1}\frac{d-1}{2}\hat a^d\vep_1^2\zt\dot\zt^2, \\
\cV_2&=M_*^{d-1}\frac{d-1}{2}\hat a^{d-2}\vep_1^2\zt\hat\pd_i\zt\hat\pd_i\zt, \\
\cV_3&=-M_*^{d-1}(d-1)\hat a^d\vep_1^2\dot\zt\hat\pd_i\zt\hat\pd^{-2}\hat\pd_i\dot\zt,
\end{split}
\label{Vi}
\end{align}
in terms of $f(\zt)$ in \eqref{f(zeta)}. We call $\cV_0$ the ``EoM vertex'' as it is proportional to the free equation of motion of $\zt$, and the rest three ``LO vertices'' as they are at the leading (second) order in the Hubble flow parameters. 
The cubic action exact in the Hubble flow parameters is given in \eqref{fullaction1} and \eqref{fullaction2}, which are the same up to total temporal and spatial derivatives.

\section{Primordial correlators by Schwinger-Keldysh path integral}

In this section, we present the method of computing equal-time correlators of the scalar fluctuation $\zt$, which is based on the path integral (or double-field) formulation of Schwinger-Keldysh's in-in formalism. After deriving the Green functions, we apply the formalism first to the power spectrum. We first derive it for the whole bulk spacetime and then obtain the part relevant to an observer localised on a 3-brane. They have already been computed in previous works~\cite{Antoniadis:2023sya,Hirose:2025pzm} in an apparently different method based on linearly perturbed Einstein equations, and we confirm that our result recovers them. The method presented in this section will be used in the next section to compute the three-point function.

\subsection{Method to compute equal-time correlators}
In this paper we compute expectation values of products of two and three scalar fluctuations $\zt$ at equal time but general spatial positions in the Bunch-Davies vacuum:
\begin{align}
\bra \zt(\hat t,\hat\bsx_1)\zt(\hat t,\hat\bsx_2) \ket, \qquad
\bra \zt(\hat t,\hat\bsx_1)\zt(\hat t,\hat\bsx_2)\zt(\hat t,\hat\bsx_3) \ket.
\end{align}
We quantise the system using the Schwinger-Keldysh in-in formalism since we are interested in expectation values defined with one state rather than transition amplitudes defined with two states at two different times. Especially, we adopt its path integral formulation, which is suitable for our cubic action for reasons we will mention later. Since the derivation of the perturbative formula for equal-time correlators is independent of spacetime dimensions, we present only the basic items to compute equal-time correlators. More details are given in for example~\cite{Palma:2023idj,Braglia:2024zsl,Kawaguchi:2024lsw,Antoniadis:2024abm}. 

Since the correlator is defined with an in-state at infinite past, the path integration variable $\zt$ lives along a path (in the complex plane of $\hat t$) that runs from infinite past to a future time and goes back to infinite past. This is equivalent to double $\zt$ to $\zt_\pm$: $\zt_+$ lives on a contour $C_+$ from the infinite past to a time $\hat t_f$ above the real axis and $\zt_-$ on a contour $C_-$ from a contour from the infinite past to a time $\hat t_f$ below the real axis:
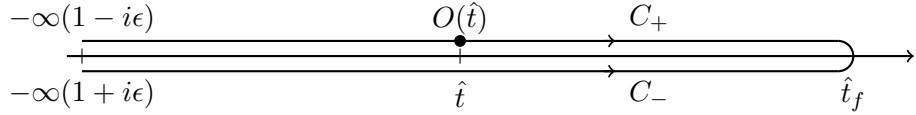
\begin{center}
\begin{tikzpicture}
    \draw[thick, ->] (-4.2, 0) -- (7, 0); 
    \draw[thick, ->] (-4, 0.2) -- (3.05, 0.2);
    \draw[thick, -] (2.9, 0.2) -- (6, 0.2);
    \draw[thick] (6, 0.2) arc (90:-90:0.2);
    \draw[thick, ->] (-4, -0.2) -- (3.05, -0.2);
    \draw[thick, -] (2.9, -0.2) -- (6, -0.2);
    \node at (6.2, -0.5) {$\hat t_f$}; 
    \node at (-4, 0.5) {$-\infty(1-i\ep)$}; 
    \node at (-4, -0.5) {$-\infty(1+i\ep)$}; 
    \draw[thin, -] (-4, -0.1) -- (-4, 0.1);
    \draw[thin, -] (1, -0.1) -- (1, 0.1);
    \node at (1, 0.5) {$O(\hat t)$}; 
    \node at (3.5, -0.5) {$C_-$};
    \node at (3.5, 0.5) {$C_+$};
    \node at (1, -0.5) {$\hat t$}; 
    \filldraw (1, 0.2) circle (2pt); 
\end{tikzpicture}
\end{center}
The $i\ep$-deviation of the paths from the real axis takes care of the Bunch-Davies vacuum.
The correlator $\bra O(\hat t) \ket$ is then given by
\begin{align}
\bra O(\hat t) \ket=\int[d\zt^\pm]\,\tilde O(\hat t)\exp\left(i\int_{C_+}\!\!\rd\hat t\rd^d\hat\bsx\,\cL^+-i\int_{C_-}\!\!\rd\hat t\rd^d\hat\bsx\,\cL^-\right),  
\end{align}
where $\cL_\pm$ is obtained by replacing $\zt$ by $\zt^\pm$ in the total Lagrangian $\cL$, and $\tilde O$ is obtained by replacing each $\zt$ in $O$ by $\zt^a$ where $a$ is $+$ or $-$, referring to the paths $C_\pm$. The choice of the path for each $\zt$ in $O$ does not affect the result: one can show that the correlator is independent of the choices of the path $a$ as long as one considers two- and three-point functions at tree level\footnote{By tree level, we mean the leading order in the expansion in $M_*^{1-d}$ in our model.} that are of our main interest.

For perturbative computations, we divide $\cL=\cL_2+\cL_I$ where $\cL_2$ is a kinetic term and $\cL_I$ is an interaction Lagrangian. One can then compute $\bra O \ket$ perturbatively based on the following  master formula:
\begin{align}
\begin{split}
\langle O(\hat t) \rangle
=\Bbra \tilde O(\hat t)\exp\left(i\int_{C_+}\rd\hat t'\rd^d\hat\bsx\,\cL_I^+(\hat t')-i\int_{C_-}\rd\hat t'\rd^d\hat\bsx\,\cL_I^-(\hat t')\right) \Kket, \label{master-O}
\end{split}
\end{align}
where $\cL_I^\pm$ is obtained by replacing $\zt$ in $\cL_I$ by $\zt^\pm$. The double bracket $\bbra\cA\kket$ denotes the Wick contraction of the operators inside $\bbra\cdots\kket$ with the following rules:
\begin{itemize}
\item Each pair $\zt^a(\hat t_1,\hat\bsx_1)\zt^b(\hat t_2,\hat\bsx_2)$ with $a,b\in\{+,-\}$ is replaced by the free Green function $G^{ab}(\hat t_1,\hat\bsx_1;\hat t_2,\hat\bsx_2)$.
\item When differential operators act on $\zt^a,\zt^b$ in a pair $\zt^a(\hat t_1,\hat\bsx_1)\zt^b(\hat t_2,\hat\bsx_2)$, one first replaces the pair $\zt^a\zt^b$ by $G^{ab}$, and then applies the differential operators to $G^{ab}$ with respect to corresponding spacetime coordinates. 
\item All $\zt^\pm$ are to be contracted. Therefore, correlators with odd numbers of $\zt^\pm$ (both from $O$ and interaction vertices) vanish. 
\end{itemize} 

An important feature of the master formula is the following: if $\cL_I$ in the master formula \eqref{master-O} takes a form of a temporal total derivative, the master formula \eqref{master-O} for $O=\zt^+\zt^+\zt^+$ with such $\cL_I$ yields zero at tree level. This justifies our use of the cubic action obtained after neglecting temporal total derivatives.\footnote{In contrast, if one uses the canonical quantisation with a Hamiltonian, the EoM vertex $\cV_0$ makes no contribution while a temporal derivative as an interaction vertex does contribute, as demonstrated in~\cite{Braglia:2024zsl,Kawaguchi:2024lsw}.} 
Another feature is that the correlator \eqref{master-O} is independent of the choice of the future endpoint of the paths $\hat t_f$  as long as it is after the time of the correlator $\hat t$. We therefore fix $\hat t_f$ to a time infinitesimally after $\hat t$, thereby almost identifying $\hat t_f$ with $\hat t$. These features are demonstrated in more details in~\cite{Braglia:2024zsl,Kawaguchi:2024lsw,Antoniadis:2024abm}.  

\subsection{Free Green functions and mode functions}

By spatial translation invariance, it is convenient to introduce the momentum space dual of the free Green functions $G^{ab}$ by
\begin{align}
\label{Gabk}
G^{ab}(\hat t,\hat\bsx;\hat t',\hat\bsx')
=\int\frac{\rd^d\hat\bsk}{(2\pi)^d}e^{i\hat\bsk\cdot(\hat\bsx-\hat\bsx')}G^{ab}_{\hat k}(\hat t,\hat t'),
\end{align}
where the momentum space Green functions depend on $\bsk$ only through its norm $\hat k=|\hat\bsk|$ due to rotation invariance. Since each direction of $\bsx_\bot$ out of $\hat\bsx$ is compactified on $S_1$ with radius $R_0$, its Fourier momentum is discretised. We denote the Fourier momentum $\hat\bsk$ as
\begin{align}
\hat\bsk=\left(\bsk,\frac{\bsn}{R_0}\right), \qquad
\bsk\in\bbR^3, \quad \bsn\in\bbZ^{d-3}\,.
\end{align} 
Therefore, the integration $\int\rd^d\hat\bsk$ should be understood as $\int\rd^3\bsk$ times the summation over $\bsn$:
\begin{align}
\label{inthatk-def}
\int\frac{\rd^d\hat\bsk}{(2\pi)^d}=\int\frac{\rd^3\bsk}{(2\pi)^3}\frac{1}{(2\pi R_0)^{d-3}}\ssum_{\bsn\in\bbZ^{d-3}} \,.
\end{align}
We will use $\int\rd^d\hat\bsk$ except when we consider the restriction of correlators on the brane.

The momentum-space free Green functions $G^{ab}_{\hat k}$ are given in terms of the momentum-space Wightman functions $G^>_{\hat k},G^<_{\hat k}$ as\footnote{
$\theta(\hat t)$ is the step function defined by $\te(\hat t)=1$ for $\hat t>0$, $\te(\hat t)=0$ for $\hat t<0$, and $\te(0)=1/2$.}
\begin{align}
G^{++}_{\hat k}(\hat t,\hat t')
&=\theta(\hat t-\hat t')G^>_{\hat k}(\hat t,\hat t')+\theta(\hat t'-\hat t)G^<_{\hat k}(\hat t,\hat t'),\\
G^{--}_{\hat k}(\hat t,\hat t')
&=\theta(t\hat '-\hat t)G^>_{\hat k}(\hat t,\hat t')+\theta(\hat t-\hat t')G^<_{\hat k}(\hat t,\hat t'), \\
G^{+-}_{\hat k}(\hat t,\hat t')&=G^<_{\hat k}(\hat t,\hat t'),\\
G^{-+}_{\hat k}(\hat t,\hat t')&=G^>_{\hat k}(\hat t,\hat t'),
\end{align}
and the momentum-space Wightman functions are given by
\begin{align}
G^{>}_{\hat k}(\hat t,\hat t')=\zt_{\hat k}(\hat t)\zt_{\hat k}^*(\hat t'), \qquad
G^{<}_{\hat k}(\hat t,\hat t')=\zt^*_{\hat k}(\hat t)\zt_{\hat k}(\hat t'),
\end{align}
where $\zt_{\hat k}$ is the mode function determined by the free equation of motion:
\begin{align}
&\pd_{\hat t}\left(\hat a(\hat t)^d\vep_1(\hat t)\pd_{\hat t}\zt_{\hat k}(\hat t)\right)
+\hat a(\hat t)^{d-2}\vep_1(\hat t)\hat k^2\zt_{\hat k}(\hat t)=0, \label{eom-mode1} \\
&(d-1)M_*^{d-1}\hat a^d\vep_1
\left(\zt_{\hat k}^*\pd_{\hat t}\zt_{\hat k}
-\zt_{\hat k}\pd_{\hat t}\zt_{\hat k}^*\right)=-i \,.
\label{eom-mode2}
\end{align}
Here $M_*$ denotes the higher dimensional reduced Planck mass.
The normalisation condition (Wronskian) in the second line above selects the Bunch-Davies vacuum. 
The Green functions then satisfy the following equations of motion:
\begin{align}
\cD_{\hat t\hat\bsx}G^{++}(\hat t,\hat\bsx;\hat t',\hat\bsx')
&=-i\de(\hat t-\hat t')\de^d(\hat\bsx-\hat\bsx'), \label{sec2:eq-G++}\\
\cD_{\hat t\hat\bsx}G^{--}(\hat t,\hat\bsx;\hat t',\hat\bsx')
&=i\de(\hat t-\hat t')\de^d(\hat\bsx-\hat\bsx'), \label{sec2:eq-G--}\\
\cD_{\hat t\hat\bsx}G^{\pm\mp}(\hat t,\hat\bsx;\hat t',\hat\bsx')&=0, \label{sec2:eq-W}
\end{align}
where $\cD_{\hat t\hat\bsx}$ is the differential operator with respect to $\hat t,\hat\bsx$:
\begin{align}\label{freeeom-D}
\cD_{\hat t\hat\bsx}f(\hat t,\hat\bsx)=
(d-1)M_*^{d-1}\left[
\pd_{\hat t}\left(\hat a(\hat t)^d\vep_1(\hat t)\pd_{\hat t}f(\hat t,\hat\bsx)\right)
-\hat a(\hat t)^{d-2}\vep_1(\hat t)\hat\pd^2f(\hat t,\hat\bsx)\right]. 
\end{align}

Let us solve \eqref{eom-mode1} and \eqref{eom-mode2}. First, we rescale $\zt_{\hat k}$ and use the conformal time $\tau$:
\begin{align}
\hat\zt_{\hat k}(\tau)=M_*^{\frac{d-1}{2}}\sqrt{(d-1)\hat a^{d-1}\vep_1}\zt_{\hat k}(\hat t(\tau))
\end{align}
which satisfies the following Klein-Gordon equation with a time-dependent mass:
\begin{align}
&\left[\pd_\tau^2+\hat k^2-\tfrac{1}{4}\hat a^2H^2(d^2-1-2(d-1)\vep_1+2d\vep_2+\vep_2^2-2\vep_1\vep_2+2\vep_2\vep_3)\right]\hat\zt_{\hat k}(\tau)=0, \\
&\hat\zt_{\hat k}(\tau)\pd_\tau\hat\zt_{\hat k}^*(\tau)-\hat\zt_{\hat k}^*(\tau)\pd_\tau\hat\zt_{\hat k}(\tau)=i\,.
\end{align}
We solve them up to the first order in the Hubble flow functions, which can then be considered time-independent (`parameters'). Up to this order, the Klein-Gordon equation reads 
\begin{align}
\left[\pd_\tau^2+\hat k^2-\tfrac{1}{4}\tau^{-2}(d^2-1+2d(d-1)\vep_1+2d\vep_2)\right]\hat\zt_{\hat k}(\tau)=0\,.
\end{align}
Its general solution can be written in terms of Hankel functions as (note that $-\hat k\tau>0$)
\begin{align}
\hat\zt_{\hat k}(\tau)=c_1(-\hat k\tau)^{1/2}H_\nu^{(1)}(-\hat k\tau)+c_2(-\hat k\tau)^{1/2}H_\nu^{(2)}(-\hat k\tau),
\end{align}
where $\nu$ is defined by
\begin{align}
\label{nu-NLO}
\nu=\frac{1}{2}\sqrt{d^2+2d(d-1)\vep_1+2d\vep_2}
=\frac{1}{2}(d+(d-1)\vep_1+\vep_2)+\cO(\vep^2)\,.
\end{align}
Since $H_\nu(z)\sim z^{-1/2}e^{iz}$ as $z\sim\infty$ (discarding inessential numerical factors), the Bunch-Davies vacuum selects $c_2=0$. The Wronskian then fixes the norm of the coefficient $c_1$ to
\begin{align}
|c_1|=\sqrt{\frac{\pi}{4k}} \,.
\end{align}
The mode function $\zt_{\hat k}$ therefore reads\footnote{Strictly speaking, this equality is up to a phase factor which does not affect the results of this paper because the correlators are written in terms of $G^>,G^<$ that are independent of the phase.}
\begin{align}
\zt_{\hat k}(\hat t)=\sqrt{\frac{\pi}{4(d-1)M_*^{d-1}}\frac{-\tau}{\vep_1\hat a^{d-1}}}H_\nu^{(1)}(-\hat k\tau)\,.
\end{align}
Therefore, the Wightman functions are given by
\begin{align}
G^{>}_{\hat k}(\hat t,\hat t') 
=\frac{\pi}{4(d-1)M_*^{d-1}}\frac{\sqrt{\tau\tau'}}{\vep_1\hat a(t)^{d-1}\hat a(t')^{d-1}}H_\nu^{(1)}(-\hat k\tau)H_\nu^{(2)}(-\hat k\tau'), \\
G^{<}_{\hat k}(\hat t,\hat t') 
=\frac{\pi}{4(d-1)M_*^{d-1}}\frac{\sqrt{\tau\tau'}}{\vep_1\hat a(t)^{d-1}\hat a(t')^{d-1}}H_\nu^{(2)}(-\hat k\tau)H_\nu^{(1)}(-\hat k\tau').
\end{align}

\subsection{Power spectrum}

Following the definition of the Green functions in momentum space \eqref{Gabk}, we define the power spectrum $\cP_{\hat k}$ by the following Fourier transformation of the two-point function\footnote{The power spectrum is usually defined in a dimensionless form $\hat k^d\cP_{\hat k}$ with a numerical factor so that it is scale invariant at the leading order in slow-roll parameters. Here we stick to the dimensionful power spectrum, as defined for general $d$ in the bulk, away from the 3-brane.}:
\begin{align}
\label{powerspectrum-def}
\bra \zt(\hat t,\hat\bsx)\zt(\hat t,\hat\bsy) \ket
=\int\frac{\rd^d\hat\bsk}{(2\pi)^d}e^{i\hat\bsk\cdot(\hat\bsx-\hat\bsy)}\cP_{\hat k}(\hat t)
\end{align}
with $\int\rd^d\hat\bsk/(2\pi)^d$ defined in \eqref{inthatk-def}.
Applying the master formula \eqref{master-O}, we find that $\cP_{\hat k}(\hat t)$ is nothing but the equal-time free Wightman function $G^>(\hat t,\hat t)$ at tree level. We therefore obtain
\begin{align}
\cP_{\hat k}(\hat t)=G^>_{\hat k}(\hat t,\hat t)=|\zt_{\hat k}(\hat t)|^2
=\frac{\pi}{4(d-1)M_*^{d-1}}\frac{-\tau}{\hat{a}^{d-1}\varepsilon_1}H^{(1)}_\nu(-\hat k\tau)H^{(2)}_\nu(-\hat k\tau)\,.
\end{align}
In the ($(d+1)$-dimensional) late-time limit $-\hat k\tau\ll1$, it asymptotes to
\begin{align}
\label{eq:Platetime}
\mathcal{P}_{\hat{k}}\simeq\frac{2^{2\nu-2}\Ga(\nu)^2}{(d-1)\pi M_*^{d-1}}\frac{(-\tau)^{1-2\nu}}{\vep_1\hat a^{d-1}}\hat k^{-2\nu},
\end{align}
where we used the asymptotic behaviour of $H^{(1)}$ as $z\sim0$:
\begin{align}\label{H1z=0}
H^{(1)}_\nu(z)\sim-\frac{2^\nu\Ga(\nu)}{\pi}z^{-\nu} \,.
\end{align}

\subsection{Power spectrum on the brane}
We are now interested in the power spectrum observed on our 3-brane, which is given by (recall the notations introduced below \eqref{metric-bg})
\begin{align}\label{<ztzt>brane}
\bra\zt(\hat t,\bsx,\bsx_\bot=\bsnl)\zt(\hat t,\bsy,\bsy_\bot=\bsnl)\ket \,.
\end{align}
Using the definition of the power spectrum \eqref{powerspectrum-def}, we can rewrite it as a Fourier transformation on the brane:
\begin{align}
\bra\zt(\hat t,\bsx,\bsnl)\zt(\hat t,\bsy,\bsnl)\ket
=\int\frac{\rd^3\bsk}{(2\pi)^3}\frac{e^{i\bsk\cdot(\bsx-\bsy)}}{(2\pi R_0)^{d-3}}\ssum_{\bsn\in\bbZ^{d-3}}\cP_{\hat k(\bsn)}, \qquad \hat k^2(\bsn)=k^2+\frac{\bsn^2}{R_0^2} \,,
\end{align}
from which we can extract the power spectrum on the brane as the sum of $\cP_{\hat k}$ over the Kaluza-Klein modes,
\begin{align}
\cP^{\mrm{brane}}_k=\frac{1}{(2\pi R_0)^{d-3}}\ssum_{\bsn\in\bbZ^{d-3}}\cP_{\hat k(\bsn)}\,.
\label{Pbrane}
\end{align}

As described in the Introduction, although higher dimensional inflation expands the size of extra dimensions relating the cosmological hierarchy of the horizon problem with the observed force of gravity, it can also generate primordial perturbations consistent with CMB observations only if the compactification size is around the micron scale which, combined with the validity of Newton's law, can work only for $d=4$ or $5$.

The $(d+1)$-dimensional inflation is then supposed to last about $\De N\sim40$ e-folds for $d=4$ (or $\sim30$ e-folds for $d=5$), while the physical momenta $\hat k/\hat aH$ decrease by a factor $e^{-40}\sim10^{-17}$ for one extra dimension (or $e^{-30}\sim10^{-13}$ for two). Therefore, almost all KK modes, including the zero mode $\hat k(\bsnl)=k$, are stretched to the superhorizon scale $-\hat k(\bsn)\tau\ll1$. Only extremely higher KK modes are within the horizon scale, but their contribution to the sum is negligible. In particular, the modes less than the (higher-dimensional) Planck scale $\hat k(\bsn)<M_*$ are all of superhorizon scale as long as $M_*\ll10^{17}$\,GeV, which covers $M_*\sim10^9$\,GeV associated with one extra dimension of micron size in the $d=4$ case (or $M_*\sim10$\,TeV for two extra dimensions of micron size in the $d=5$ case), as explained in the Introduction. Therefore, we may take the sum over the KK modes after taking the late time limit of $\cP_{\hat k}$. 
Substituting \eqref{eq:Platetime} into \eqref{Pbrane}, we obtain the power spectrum on the brane at late time,
\begin{align}
\cP^{\mrm{brane}}_k\simeq
\frac{2^{2\nu-2}\Ga(\nu)^2R_0^{2\nu}}{(d-1)\pi(2\pi R_0)^{d-3}M_*^{d-1}}
\frac{(-\tau)^{1-2\nu}}{\vep_1\hat a^{d-1}}S_{\nu}(R_0^2k^2)\,,
\end{align}
where we introduced the infinite sum $S_\alpha(z)$ over $\bsn$:
\begin{align}\label{Salpha}
    S_\alpha(z):=\sum\limits_{\bsn\in\bbZ^{d-3}}\frac{1}{\big(z+\bsn^2\big)^\alpha}\,.
\end{align}
Let us look at two limits $R_0k\ll1$ and $R_0k\gg1$, in which the sum $S_\alpha$ behaves as~\cite{Antoniadis:2023sya, Hirose:2025pzm}
\begin{align}
\begin{split}
    S_\alpha(z)&\underset{z\ll1}{\simeq}\frac{1}{z^\alpha}, \\
    S_\alpha(z)&\underset{z\gg1}{\simeq}\frac{\pi^{\frac{d-3}{2}}\Gamma\left(\alpha-\frac{d-3}{2}\right)}{\Gamma\left(\alpha\right)}z^{\frac{d-3}{2}-\alpha},
\end{split} \label{Salpha_limits}
\end{align}
where the first line means that in the case $z\ll1$, the contributions from all nonzero KK modes are negligible and only the zero mode survives, whereas the second line exhibits a different scaling in $z$ due to the summation over KK modes. Therefore, the power spectrum on the brane in the two limits behaves as
\begin{align}
\cP^{\mrm{brane}}_k&\underset{R_0k\ll1}{\simeq}
\frac{2^{2\nu-2}\Ga(\nu)^2}{(d-1)\pi(2\pi R_0)^{d-3}M_*^{d-1}}
\frac{(-\tau)^{1-2\nu}}{\vep_1\hat a^{d-1}k^{2\nu}}, \\
\cP^{\mrm{brane}}_k&\underset{R_0k\gg1}{\simeq}
\frac{2^{2\nu-2}\Ga(\nu)\Ga(\nu-\frac{d-3}{2})}{(d-1)\pi^{\frac{5-d}{2}}(2\pi)^{d-3}M_*^{d-1}}
\frac{(-\tau)^{1-2\nu}}{\vep_1\hat a^{d-1}k^{2\nu-d+3}}\,.
\end{align}
This result recovers the 4D ($d=3$) scalar power spectrum, the 5D ($d=4$) one~\cite{Antoniadis:2023sya}, and the one for general $d$~\cite{Hirose:2025pzm}.

The scalar spectral index $n_s$ is read off from the dimensionless version of the power spectrum 
$P_k=\cP_k^\text{brane}k^3/(2\pi^2)$:
\begin{align}
\label{Pk}
P_k=A_s\left(\frac{k}{k_*}\right)^{n_s-1},
\end{align}
where $k_*$ is a pivot momentum and $A_s$ is the amplitude defined as the power spectrum for $k=k_*$. The spectral indices of the 3-brane power spectrum in the two limits then read
\begin{align}
    n_s-1&\underset{R_0k\ll1}{\simeq}3-2\nu\simeq3-d-(d-1)\varepsilon_1-\varepsilon_2,\\
    n_s-1&\underset{R_0k\gg1}{\simeq}d-2\nu\simeq-(d-1)\varepsilon_1-\varepsilon_2,\label{eq:spectralindexsmall}
\end{align}
where we used \eqref{nu-NLO}. One can see that (approximate) scale invariance is recovered when $R_0k\gg1$, namely when the 3D comoving wavelength is smaller than the compactification size, which is a consequence of taking the sum over KK modes in $S_\nu$ as commented above. On the other hand, when the comoving wavelength is much larger than the compactification size, the spectral index exhibits $\cO(1)$ violation of scale invariance.

\section{Bispectrum}

Let us move on to the bispectrum, which is defined in momentum space by
\begin{align}
\bra\zt(\hat{t},\hat\bsx_1)\zt(\hat{t},\hat\bsx_2)\zt(\hat{t},\hat\bsx_3)\ket
&=\int_{\hat\bsk_1,\hat\bsk_2,\hat\bsk_3}
e^{i\sum_{a=1}^3\hat\bsk_a\cdot\hat\bsx_a}(2\pi)^d\delta^{d}\left(\ssum_{a=1}^3\hat\bsk_a\right)\cB_{\hat k_1,\hat k_2,\hat k_3}(\hat t),
\end{align}
where we introduced the shorthand notation
\begin{align}
\int_{\hat\bsk_1,\hat\bsk_2,\hat\bsk_3}:=\int\frac{\rd^d\hat\bsk_1}{(2\pi)^d}\frac{\rd^d\hat\bsk_2}{(2\pi)^d}\frac{\rd^d\hat\bsk_3}{(2\pi)^d}\,.
\end{align}
We are interested in this at tree level and leading order in Hubble flow parameters. Therefore, the relevant cubic vertices are $\cV_0,\cV_1,\cV_2,\cV_3$ in \eqref{Vi}. Applying the master formula \eqref{master-O} with these vertices, we can write $\cB$ as a sum of four contributions:
\begin{align} \label{fourI}
\cB_{\hat k_1,\hat k_2,\hat k_3}(\hat t)
&=I_0(\hat t)+I_1(\hat t)+I_2(\hat t)+I_3(\hat t),
\end{align}
where $I_i$ is the result of applying the master formula \eqref{master-O} with $\cL_I=\cV_i$ in \eqref{Vi}. 
In order to compute them, we need the time derivatives of the Wightman functions defined by
\begin{align}
\dot{G}_{\hat{k}}^>\big(\hat{t},\hat{t}'\big):=\zeta_{\hat{k}}\big(\hat{t}\big)\dot{\bar{\zeta}}_{\hat{k}}\big(\hat{t}'\big), \qquad
\dot{G}_{\hat{k}}^<\big(\hat{t},\hat{t}'\big):=\bar{\zeta}_{\hat{k}}\big(\hat{t}\big)\dot{\zeta}_{\hat{k}}\big(\hat{t}'\big).
\end{align}
At the leading order in Hubble flow parameters, the mode function is approximated to
\begin{align}
\zt_{\hat k}(\hat t)=\sqrt{C}(-\tau)^{d/2}H_{d/2}^{(1)}(-\hat k\tau)+\cO(\vep^0),
\end{align}
where we introduced the parameter
\begin{align}
C:=\frac{\pi H^{d-1}}{4(d-1)M_*^{d-1}\vep_1} \label{Cdef}\,.
\end{align}
In terms of this, the Wightman functions and their time derivatives read
\begin{align}
G^>_{\hat k}(\hat t,\hat t')&=C(\tau\tau')^{d/2}H_{d/2}^{(1)}(-\hat k\tau)H_{d/2}^{(2)}(-\hat k\tau')+\cO(\vep^0), \\
{\dot G}^>_{\hat k}(\hat t,\hat t')&=CH\hat k\tau'(\tau\tau')^{d/2}H_{d/2}^{(1)}(-\hat k\tau)H_{(d/2)-1}^{(2)}(-\hat k\tau')+\cO(\vep^0),
\end{align}
and
\begin{align}
G^<_{\hat k}(\hat t,\hat t')&=C(\tau\tau')^{d/2}H_{d/2}^{(2)}(-\hat k\tau)H_{d/2}^{(1)}(-\hat k\tau')+\cO(\vep^0), \\
{\dot G}^<_{\hat k}(\hat t,\hat t')&=CH\hat k\tau'(\tau\tau')^{d/2}H_{d/2}^{(2)}(-\hat k\tau)H_{(d/2)-1}^{(1)}(-\hat k\tau')+\cO(\vep^0)\,. \label{G<dot}
\end{align}
At late time $G^>_{\hat k}$ asymptotes to a constant in $\tau$ while ${\dot G}^>_{\hat k}$ becomes order $(-\hat k\tau)^2$. Therefore ${\dot G}^>_{\hat k}$ becomes subleading at late time compared with $G^>_{\hat k}$.

In the following, we will set $M_*=1$.

\subsection{EoM vertex}

Let us first look at the factor $I_0$ from the EoM vertex $\cV_0$.
In the path integral formulation, this vertex does contribute to the bispectrum as demonstrated in~\cite{Braglia:2024zsl,Kawaguchi:2024lsw,Antoniadis:2024abm}. Wick contractions of three $\zt^+$ and the vertex $\cL_I=\cV_0$ yields two free Green functions $G^{+\pm}$ and one with the action of the differential operator $\cD G^{+\pm}$ where $\cD$ is given in \eqref{freeeom-D}. Since $\cD G^{++}$ is equal to a delta function as in \eqref{sec2:eq-G++} and $\cD G^{+-}$ vanishes as in \eqref{sec2:eq-W}, the result is given in terms of the equal-time Wightman function $G^>(\hat t,\hat t)$ as
\begin{align}
I_0(\hat t)=\frac{\vep_2}{2}\left[
G^>_{\hat k_1}(\hat t,\hat t)G^>_{\hat k_2}(\hat t,\hat t)
+G^>_{\hat k_1}(\hat t,\hat t)G^>_{\hat k_3}(\hat t,\hat t)
+G^>_{\hat k_2}(\hat t,\hat t)G^>_{\hat k_3}(\hat t,\hat t)
\right]+\cO(\vep^0,-\hat k\tau),
\end{align}
where we only picked up the leading order terms in $M_*^{1-d}$ (or equivalently loop) expansion. Moreover, we only kept the factors at leading order in the late time limit $-k\tau\to0$. This selects the terms coming from the Wick contractions with $\vep_2\zt^2/4$ out of $f(\zt)$ in $\cV_0$ since the other terms in $f(\zt)$ contain time and space derivatives and are hence subdominant in the late time limit (see comment below \eqref{G<dot}).
Using the asymptotic form of $H^{(1)}$ \eqref{H1z=0}, we obtain the late time limit of $I_0$:
\begin{align}
I_0(\hat t)=\frac{2^{2d-5}\Ga(\frac{d}{2})^4H^{2d-2}\vep_2}{\pi^2(d-1)^2\vep_1^2(\hat k_1\hat k_2\hat k_3)^d}(\hat k_1^d+\hat k_2^d+\hat k_3^d)+\cO(-\hat k\tau)\,.
\end{align}

\subsection{LO vertices}
Next, $I_1,I_2,I_3$ from the LO vertices in~\eqref{fourI} are given at the leading order by
\begin{align}
\begin{split} 
I_1(\hat{t})&=i\frac{(d-1)\varepsilon_1^2}{H^{d+1}}
\Big(I^\bullet_{\hat{k}_1\hat{k}_2\hat{k}_3}+I^\bullet_{\hat{k}_2\hat{k}_3\hat{k}_1}+I^\bullet_{\hat{k}_3\hat{k}_1\hat{k}_2}\Big),
\end{split}\label{I1def}\\
\begin{split}
I_2\big(\hat{t}\big)&=i\frac{(d-1)\varepsilon_1^2}{H^{d-1}}\frac{\hat{k}_1^2+\hat{k}_2^2+\hat{k}_3^2}{2}I_{\hat{k}_1\hat{k}_2\hat{k}_3},
\end{split}\label{I2def}\\
\begin{split} 
I_3\big(\hat{t}\big)&=-i\frac{(d-1)\varepsilon_1^2}{H^{d+1}}
\Bigg\{\left(\frac{\hat{k}_2^2-\hat{k}_3^2-\hat{k}_1^2}{2\hat{k}_3^2}+\frac{\hat{k}_3^2-\hat{k}_2^2-\hat{k}_1^2}{2\hat{k}_2^2}\right)I^\bullet_{\hat{k}_1\hat{k}_2\hat{k}_3}\\
&\qquad\qquad\qquad\quad\ \ \ \ 
+\left(\frac{\hat{k}_1^2-\hat{k}_3^2-\hat{k}_2^2}{2\hat{k}_3^2}+\frac{\hat{k}_3^2-\hat{k}_1^2-\hat{k}_2^2}{2\hat{k}_1^2}\right)I^\bullet_{\hat{k}_2\hat{k}_3\hat{k}_1}\\
&\qquad\qquad\qquad\quad\ \ \ \ 
+\left(\frac{\hat{k}_2^2-\hat{k}_1^2-\hat{k}_3^2}{2\hat{k}_1^2}+\frac{\hat{k}_1^2-\hat{k}_2^2-\hat{k}_3^2}{2\hat{k}_2^2}\right)I^\bullet_{\hat{k}_3\hat{k}_1\hat{k}_2}\Bigg\}\,,
\end{split}
\label{I3def}
\end{align}
where we introduced the following integrals:
\begin{align}
I_{\hat{k}_1\hat{k}_2\hat{k}_3}&:=
\int_{\tau'\in c_1} \!\! G^>_{\hat{k}_1}\big(\hat{t},\hat{t}'\big)G^>_{\hat{k}_2}\big(\hat{t},\hat{t}'\big)G^>_{\hat{k}_3}\big(\hat{t},\hat{t}'\big)
-\int_{\tau'\in\ol{c_1}} \!\! G^<_{\hat{k}_1}\big(\hat{t},\hat{t}'\big)G^<_{\hat{k}_2}\big(\hat{t},\hat{t}'\big)G^<_{\hat{k}_3}\big(\hat{t},\hat{t}'\big),\\
I^\bullet_{\hat{k}_1\hat{k}_2\hat{k}_3}&:=
\int_{\tau'\in c_1} \!\! \frac{1}{\tau^{'2}}G^>_{\hat{k}_1}\big(\hat{t},\hat{t}'\big)\dot{G}^>_{\hat{k}_2}\big(\hat{t},\hat{t}'\big)\dot{G}^>_{\hat{k}_3}\big(\hat{t},\hat{t}'\big)
-\int_{\tau'\in\ol{c_1}} \! \frac{1}{\tau^{'2}}G^<_{\hat{k}_1}\big(\hat{t},\hat{t}'\big)\dot{G}^<_{\hat{k}_2}\big(\hat{t},\hat{t}'\big)\dot{G}^<_{\hat{k}_3}\big(\hat{t},\hat{t}'\big),
\end{align}
with the integral symbols $\int_{\tau'\in c_1}$ and $\int_{\tau'\in\ol{c_1}}$ over $\tau'$ defined by 
\begin{align}
\int_{\tau'\in c_1}:=\int_{c_1}\frac{\rd\tau'}{(-\tau')^{d-1}}, \qquad
\int_{\tau'\in\ol{c_1}}:=\int_{\ol{c_1}}\frac{\rd\tau'}{(-\tau')^{d-1}}\,.
\end{align}
The contour $c_1$ runs from $-\infty(1-i\ep)$ to $\tau$ and the contour $\ol{c_1}$ runs from $-\infty(1+i\ep)$ to $\tau$, as depicted in Figure~\ref{fig:taucomplexplane}.

\begin{figure}[!htb]
   \begin{minipage}{0.48\textwidth}
     \begin{center}
\begin{tikzpicture}
    \draw[->] (-3,0) -- (3,0) node[right] {};
    \draw[->] (0,-3) node[below]{$\phantom{\big(\big)} $} -- (0,3) node[above] {$\phantom{\big(\big)} $};
    
    \draw (-0.25,-0.1) -- (-0.25,0.1) node[above] {$\tau$};

     \draw[->, blue, thick] (-3,0.25) node[above] {$-\infty(1-i\epsilon)$} -- (-1.625,0.125) node[above]{$c_1$};
\draw[-, blue, thick] (-1.625,0.125) -- (-0.25,0);

\draw[->, red, thick] (-3,-0.25) node[below]  {$-\infty(1+i\epsilon)$}--(-1.625,-0.125) node[below]{$\ol{c_1}$};
\draw[-, red, thick]  (-1.625,-0.125)--(-0.25,0) ;
\end{tikzpicture}
\caption{$\tau'$ complex plane
}
\label{fig:taucomplexplane}
\end{center}
   \end{minipage}\hfill
   \begin{minipage}{0.48\textwidth}
    \begin{center}
\begin{tikzpicture}
    \draw[->] (-3,0) -- (3,0) node[right] {};
    \draw[->] (0,-3) -- (0,3) node[above] {};
    
    \draw[-] (-0.1,0.25)  -- (0.1,0.25) node[left] {$-i\tau\ $};
    \draw (-0.1,-0.25) -- (0.1,-0.25) node[left] {$i\tau\ $};
    \draw (0.25,-0.1) -- (0.25,0.1) node[above right] {};

     \draw[->, blue, thick] (0.25,3) node[above] {$i\infty(1-i\epsilon)$} -- (0.125,1.625) node[right]{$\gamma_1$};
\draw[-, blue, thick] (0.125,1.625) -- (0,0.25);
\draw [->, blue, thick, dashed] (0,0.25) arc[start angle=90, end angle=45, radius=0.25] node[above right]{$\gamma_2$};
\draw [-, blue, thick, dashed] (0.176777,0.176777) arc[start angle=45, end angle=0, radius=0.25];
\draw [->, blue, thick, dashed] (3,0) arc[start angle=0, end angle=45, radius=3] ;
\draw [-, blue, thick, dashed] (2.12132,2.12132) arc[start angle=45, end angle=85, radius=3];

\draw[->, red, thick] (0.25,-3) node[below]  {$-i\infty(1+i\epsilon)$}--(0.125,-1.625) node[right]{$\overline{\gamma_1}$} ;
\draw[-, red, thick]  (0.125,-1.625)--(0,-0.25) ;
\draw [->, red, thick, dashed] (0,-0.25) arc[start angle=-90, end angle=-45, radius=0.25]  node[below right]{$\overline{\gamma_2}$};
\draw [-, red, thick, dashed] (0.176777,-0.176777) arc[start angle=-45, end angle=0, radius=0.25];
\draw [->, red, thick, dashed] (3,0) arc[start angle=0, end angle=-45, radius=3];
\draw [-, red, thick, dashed] (2.12132,-2.12132) arc[start angle=-45, end angle=-85, radius=3];

\draw[->, purple, thick, dashed] (0.25,0) --(1.375,0) ;
\draw[-, purple, thick, dashed]  (1.375,0)--(3,0) node[right] {$\infty$} ;
\end{tikzpicture}
\caption{$x$ complex plane
}
\label{fig:xcomplexplane}
\end{center}
   \end{minipage}
\end{figure}

We now make two changes of variable: $\tau'=ix$ for $\tau'$ along $c_1$ and $\tau'=-ix$ for $\tau'$ along $\ol{c_1}$, which rotate $c_1,\ol{c_1}$ to contours $\ga_1,\ol{\ga_1}$ in Figure~\ref{fig:xcomplexplane}, respectively. These rotations transform the Hankel function into the deformed Bessel function $K_\nu$:
\begin{align}
{\displaystyle K_{\nu}(z)=
{\begin{cases}
{\frac {\pi }{2}}i^{\nu +1}H_{\nu }^{(1)}(iz)&-\pi <\arg z\leq {\frac {\pi }{2}}, \\
{\frac {\pi }{2}}(-i)^{\nu +1}H_{\nu }^{(2)}(-iz)&-{\frac {\pi }{2}}<\arg z\leq \pi. \end{cases}}}
\end{align} 
The integral $I_{\hat{k}_1\hat{k}_2\hat{k}_3}$ then becomes
\begin{align}
I_{\hat{k}_1\hat{k}_2\hat{k}_3}=
i\frac{8C^3}{\pi^3}
\left[
i^{d+2}\cH^{(1)}(\tau)\cI_{\gamma_1}^{(\frac{d}{2}+1,\frac{d}{2},\frac{d}{2},\frac{d}{2})}
+(-i)^{d+2}\cH^{(2)}(\tau)\cI_{\overline{\gamma_1}}^{(\frac{d}{2}+1,\frac{d}{2},\frac{d}{2},\frac{d}{2})}
\right],
\label{Ik1k2k3-gamma1}
\end{align}
where we have introduced the shorthand notation
\begin{align}
\cH^{(a)}(\tau):=(-\tau)^{\frac{3d}{2}}H^{(a)}_{d/2}(-\hat{k}_1\tau)H^{(a)}_{d/2}(-\hat{k}_2\tau)H^{(a)}_{d/2}(-\hat{k}_3\tau) \quad (a=1,2),
\end{align}
and the triple $K_\nu$ integral along contour $\cC$:
\begin{align}
\cI_{\mathcal{C}}^{(\alpha,\beta_1,\beta_2,\beta_3)}:=\int_\mathcal{C}dx\ x^\alpha K_{\beta_1}(\hat{k}_1x)K_{\beta_2}(\hat{k}_2x)K_{\beta_3}(\hat{k}_3x).
\end{align}
By Cauchy's theorem, one can deform $\ga_1$ into the contour $\ga_2:-i\tau\to-\tau$ plus the line $[-\tau,\infty)$ plus the arc $\infty\to i\infty(1-i\ep)$ with infinitely large radius, as depicted in Figure~\ref{fig:xcomplexplane}. However, the integral $\cI_\cC$ along the last one (the big arc) is zero because $K_\nu$ damps exponentially around infinity. We therefore obtain
\begin{align}
\cI_{\gamma_1}^{(\frac{d}{2}+1,\frac{d}{2},\frac{d}{2},\frac{d}{2})}=
-\cI_{[-\tau,\infty)}^{(\frac{d}{2}+1,\frac{d}{2},\frac{d}{2},\frac{d}{2})}-\cI_{\gamma_2}^{(\frac{d}{2}+1,\frac{d}{2},\frac{d}{2},\frac{d}{2})}\,. 
\label{cI_ga1}
\end{align}
The integral $\cI_{[-\tau,\infty)}$ over real numbers in $[-\tau,\infty)$ is finite because it has a finite cutoff $-\tau$ near the origin and $K_\nu(z)$ decays exponentially at large $z$.
Substituting \eqref{cI_ga1} back into \eqref{Ik1k2k3-gamma1}, we obtain
\begin{align}
\begin{split}
I_{\hat{k}_1\hat{k}_2\hat{k}_3}
=-i\frac{16C^3}{\pi^3}
\,\text{Re}\left\{i^{d+2}\cH^{(1)}(\tau)
\left[\cI_{[-\tau,\infty)}^{(\frac{d}{2}+1,\frac{d}{2},\frac{d}{2},\frac{d}{2})}+\cI_{\gamma_2}^{(\frac{d}{2}+1,\frac{d}{2},\frac{d}{2},\frac{d}{2})}\right]\right\}\,.
\end{split}
\label{eq:Itransformed}
\end{align}
Applying the same prescription to $I^\bullet_{\hat{k}_1\hat{k}_2\hat{k}_3}$ gives
\begin{align}
\begin{split}\label{eq:Idottransformed}
I^\bullet_{\hat{k}_1\hat{k}_2\hat{k}_3}
=-i\frac{16C^3H^2}{\pi^3}\hat k_2\hat k_3
\,\text{Re}\left\{i^d\cH^{(1)}(\tau)
\left[\cI_{[-\tau,\infty)}^{(\frac{d}{2}+1,\frac{d}{2},\frac{d}{2}-1,\frac{d}{2}-1)}+\cI_{\gamma_2}^{(\frac{d}{2}+1,\frac{d}{2},\frac{d}{2}-1,\frac{d}{2}-1)}\right]\right\}\,.
\end{split}
\end{align}

\paragraph{Late time limit}
Let us now compute \eqref{eq:Itransformed} and \eqref{eq:Idottransformed} in the late time limit $-\hat k\tau\to0$.
In the following arguments, taking the real part is crucial. The behaviour of the integrals then changes a lot depending on whether $d$ is odd or even due to the factors $i^{d+2},i^d$. 

Before proceeding, we present two properties of the Bessel functions $H_\nu,K_\nu$ and their integrals that will play crucial role:
\begin{enumerate}[label=(\arabic*)]
\item 
Hankel function $H^{(1)}_\nu$ is defined as $H^{(1)}_\nu=J_\nu+iN_\nu$, where Bessel $J_\nu(z)$ and Neumann $N_\nu(z)$ functions are real-valued as long as $\nu$ and $z$ are positive, and near $z\sim 0$ they behave as $J_\nu(z)\sim z^\nu[1+\cO(z^2)]$ and $N_\nu(z)\sim z^{-\nu}[1+\cO(z^2)]$. Therefore, when $H^{(1)}_\nu$ is expanded in small $z$, the coefficients of powers from $z^{-\nu}$ to $z^{\nu-1}$ are pure imaginary and the real coefficients appear at powers $z^\nu$ and higher. 

\item
Let us summarise now the properties of the integrals $\cI_{[-\tau,\infty)}$ in \eqref{eq:Itransformed} and \eqref{eq:Idottransformed}. First, they are real because so is $K_\nu(z)$ for positive $\nu$ and $z$. In general it is difficult to find an analytic expression of $\cI_{[-\tau,\infty)}$ as it is an indefinite integral. However, the steps of taking the real part and the late time limit facilitate the computations. The late time behavour can be seen as follows: one first decomposes the integration region into $[-\tau,\al]$ and $[\al,\infty)$, where a number $\al>0$ is chosen so small that each $K_\nu$ in the integral can be approximated by its leading term of order $z^{-\nu}$. Then $\cI_{[\al,\infty)}$ just gives a number independent of $\tau$ while the rest $\cI_{[-\tau,\al]}$ at late time can be estimated by picking up the leading term of each $K_\nu$.
Using 
\begin{align}\label{Ksmallz}
K_{\nu}(z)\underset{z\rightarrow0}{\sim}\frac{1}{2}\Gamma\left(\nu\right)\left(\frac{z}{2}\right)^{-\nu},
\end{align}
we obtain
\begin{align}\label{eq:Ireal}
\cI_{[-\tau,\infty)}^{(\alpha,\beta_1,\beta_2,\beta_3)}\underset{\tau\rightarrow0^-}{\sim}
\frac{2^{\beta_1+\beta_2+\beta_3-3}\Gamma(\beta_1)\Gamma(\beta_2)\Gamma(\beta_3)}{1+\alpha-\beta_1-\beta_2-\beta_3}\frac{(-\tau)^{1+\alpha-\beta_1-\beta_2-\beta_3}}{\hat{k}_1^{\beta_1}\hat{k}_2^{\beta_2}\hat{k}_3^{\beta_3}},
\end{align}
where we suppressed the constants from $\cI_{[\al,\infty)}$ and the upper end $\al$ of the integral $\cI_{[-\tau,\al]}$. In particular, the integrals in \eqref{eq:Itransformed} and \eqref{eq:Idottransformed} behave as
\begin{align}
&\cI_{[-\tau,\infty)}^{(\frac{d}{2}+1,\frac{d}{2},\frac{d}{2},\frac{d}{2})} \sim (-\tau)^{2-d}, \label{cI1-leading} \\
&\cI_{[-\tau,\infty)}^{(\frac{d}{2}+1,\frac{d}{2},\frac{d}{2}-1,\frac{d}{2}-1)} \sim (-\tau)^{4-d} \label{cI2-leading}\,.
\end{align}
There are however some caveats on \eqref{cI2-leading} for $d=3,4$. For $d=3$, the leading term of \eqref{cI2-leading} at late time is apparently $(-\tau)^1$, but it is wrong since we have neglected constant contributions as already mentioned above. Indeed, the correct leading term is of order $(-\tau)^0$, but its value is not given by \eqref{eq:Ireal}. For $d=4$, the leading term of \eqref{cI2-leading} is correctly at the zeroth order, but its value is not necessarily given by \eqref{eq:Ireal} for the same reason.

\end{enumerate}

\subsubsection{Odd $d$}
Let us first look at the contribution from the terms with $\cI_{[-\tau,\infty)}$ in \eqref{eq:Itransformed} and \eqref{eq:Idottransformed}. Since these integrals are real and $i^d$ is pure imaginary, we need the imaginary part of $\cH^{(1)}$, which, according to the property (1) above, is of order constant:
\begin{align}
\text{Im}[\cH^{(1)}(\tau)]=\cO((-\tau)^0).
\end{align}
Therefore we need the following terms from the integrals in \eqref{eq:Itransformed} and \eqref{eq:Idottransformed}:
\begin{align}
\cI_{[-\tau,\infty)}^{(\frac{d}{2}+1,\frac{d}{2},\frac{d}{2},\frac{d}{2})}: \quad (-\tau)^{2-d}, \cdots, (-\tau)^{-1}, (-\tau)^0, \\
\cI_{[-\tau,\infty)}^{(\frac{d}{2}+1,\frac{d}{2},\frac{d}{2}-1,\frac{d}{2}-1)}: \quad (-\tau)^{4-d}, \cdots, (-\tau)^{-1}, (-\tau)^0.
\end{align}
For $d=3$, we need the $(-\tau)^{-1}$ and $(-\tau)^0$ terms. As already explained above as a caveat, the constant contribution $(-\tau)^0$ cannot be obtained by \eqref{eq:Ireal}. Fortunately, we can still compute the integral as $K_\nu(z)$ with half integer $\nu$ can be rewritten as a product of a polynomial times $z^{-\nu}e^{-z}$. 
For higher odd $d$, we need terms with higher negative powers of $-\tau$ with exact coefficients as well as the $(-\tau)^0$ terms, for which we use the explicit expression of $K_\nu$ for half integer $\nu$ in terms of elementary functions.

Let us consider the case $d=3$. In this case, $\cH^{(1)}$ is expanded as
\begin{align}
&\cH^{(1)}(\tau)|_{d=3}
=\frac{i\sqrt{2}}{\pi^\frac{3}{2}(\hat{k}_1\hat{k}_2\hat{k}_3)^\frac{3}{2}}\left(2+\big(\hat{k}_1^2+\hat{k}_2^2+\hat{k}_3^2\big)\tau^2-\frac{2i}{3}\big(\hat{k}_1^3+\hat{k}_2^3+\hat{k}_3^3\big)\tau^3\right)+\mathcal{O}\big(\tau^4\big)\,.
\end{align}
The integrals $\cI_{[-\tau,\infty)}$ can be computed exactly by using $K_{3/2}(z)=\sqrt{\pi/2}z^{-3/2}(1+z)e^{-z}$. We then obtain
\begin{align}
\text{Re}\left[i^{5}\cH^{(1)}(\tau)
\cI_{[-\tau,\infty)}^{(\frac{5}{2},\frac{3}{2},\frac{3}{2},\frac{3}{2})}\right]
&=\frac{1}{\hat{k}_1^3 \hat{k}_2^3 \hat{k}_3^3 }\left(\frac{1}{\tau} + \hat{k}_t - \frac{\hat{k}_1 \hat{k}_2 + \hat{k}_1 \hat{k}_3 + \hat{k}_2 \hat{k}_3}{\hat{k}_t} - \frac{\hat{k}_1 \hat{k}_2 \hat{k}_3}{\hat{k}_t^2} \right)+\mathcal{O}(\tau),
\label{line_I1d3} \\
\text{Re}\left[i^{3}\cH^{(1)}(\tau)
\cI_{[-\tau,\infty)}^{(\frac{5}{2},\frac{3}{2},\frac{1}{2},\frac{1}{2})}\right]
&=\frac{2\hat{k}_1+\hat{k}_2+\hat{k}_3}{\hat{k}_1^3\hat{k}_2\hat{k}_3\hat{k}_t^2}+\mathcal{O}(\tau)\,,
\label{line_I2d3}
\end{align}
where we introduced $\hat k_t=\hat k_1+\hat k_2+\hat k_3$.

Let us proceed to the part with $\cI_{\ga_2}$. Since the radius of $\ga_2$ is $-\tau$ and the late time limit $-\tau\to0$ is taken, it amounts to computing the angular integral
\begin{align}\label{angleint}
\cI_{\gamma_2}^{(\alpha,\beta_1,\beta_2,\beta_3)}=i\int_{\frac{\pi}{2}}^{0}d\theta\ (-\tau e^{i\theta})^{\alpha+1} K_{\beta_1}(-\hat{k}_1\tau e^{i\theta})K_{\beta_2}(-\hat{k}_2\tau e^{i\theta})K_{\beta_3}(-\hat{k}_3\tau e^{i\theta})
\end{align}
with the asymptotic form \eqref{Ksmallz}. For example, when $d=3$, the result is
\begin{align}
\text{Re}\left[i^{5}\cH^{(1)}(\tau)\cI_{\gamma_2}^{(\frac{5}{2},\frac{3}{2},\frac{3}{2},\frac{3}{2})}\right]
&=-\frac{1}{\hat{k}_1^3 \hat{k}_2^3 \hat{k}_3^3 \tau} +\mathcal{O}(\tau), \label{ga2_I1d3} \\
\text{Re}\left[i^{3}\cH^{(1)}(\tau)\cI_{\gamma_2}^{(\frac{5}{2},\frac{3}{2},\frac{1}{2},\frac{1}{2})}\right]
&=\mathcal{O}(\tau). \label{ga2_I2d3}
\end{align}
Notice that the $1/\tau$ term in \eqref{ga2_I1d3} cancels that in \eqref{line_I1d3}.

Summing up all contributions and using $C$ defined by \eqref{Cdef}, we obtain
\begin{align}
I_{\hat{k}_1\hat{k}_2\hat{k}_3}&=
-\frac{iH^{6}}{32\varepsilon_1^3\hat{k}_1^3 \hat{k}_2^3 \hat{k}_3^3} \left( \hat{k}_t - \frac{\hat{k}_1 \hat{k}_2 + \hat{k}_1 \hat{k}_3 + \hat{k}_2 \hat{k}_3}{\hat{k}_t} - \frac{\hat{k}_1 \hat{k}_2 \hat{k}_3}{\hat{k}_t^2} \right)+\mathcal{O}(\tau)\,,\\
I^\bullet_{\hat{k}_1\hat{k}_2\hat{k}_3}&=
-\frac{iH^{8}(2\hat{k}_1 + \hat{k}_2 + \hat{k}_3)}{32\varepsilon_1^3\hat{k}_1^3 \hat{k}_2 \hat{k}_3\hat{k}_t^2}+\mathcal{O}(\tau)\,.
\end{align}
Plugging them back into \eqref{I1def}--\eqref{I3def}, we obtain $I_1(\hat t),I_2(\hat t),I_3(\hat t)$:
\begin{align}
\begin{split}
I_1\big(\hat{t}\big)|_{d=3}&=
\frac{H^4}{16\varepsilon_1\hat{k}_1^3\hat{k}_2^3\hat{k}_3^3}\ssum_{1\leq a<b \leq 3}\left(\frac{\hat{k}_a^2\hat{k}_b^2}{\hat k_t}
+\frac{\hat{k}_1\hat{k}_2\hat{k}_3}{\hat k_t^2}\hat{k}_a\hat{k}_b\right)+\mathcal{O}\big(\tau,\varepsilon^0\big),
\end{split} \\
\begin{split}
I_2\big(\hat{t}\big)|_{d=3}&=
\frac{H^4}{32\varepsilon_1\hat{k}_1^3\hat{k}_2^3\hat{k}_3^3}\left(\ssum_{a=1}^3\hat{k}_a^2\right)
\bigg(\hat{k}_t-\frac{\hat{k}_1\hat{k}_2\hat{k}_3}{\hat{k}_t^2}
-\frac{1}{\hat{k}_t}\ssum_{1\leq a<b \leq 3}\hat{k}_a\hat{k}_b\bigg)+\mathcal{O}\big(\tau,\varepsilon^0\big),
\end{split} \\
\begin{split}
I_3\big(\hat{t}\big)|_{d=3}&=\frac{H^4}{32\varepsilon_1\hat{k}_1^3\hat{k}_2^3\hat{k}_3^3}
\left[
\ssum_{a=1}^3\bigg(\frac{2\hat{k}_1\hat{k}_2\hat{k}_3}{\hat{k}_t^2}\hat{k}_a^2-\hat{k}_a^3-\frac{\hat{k}_a^4}{\hat k_t}\bigg)
+\frac{6}{\hat k_t}\ssum_{1\leq a<b\leq 3}\hat{k}_a^2\hat{k}_b^2
\right]+\mathcal{O}\big(\tau,\varepsilon^0\big)\,.
\end{split}
\end{align}
Finally, summing up $I_0,I_1,I_2,I_3$, we obtain
\begin{align}
\mathcal{B}_{\hat k_1,\hat k_2,\hat k_3}^{d=3}
&=\frac{H^4}{32\varepsilon_1^2\hat{k}_1^3\hat{k}_2^3\hat{k}_3^3}
\left[
(\varepsilon_2-2\varepsilon_1)\ssum_{a=1}^3\hat{k}_a^3
+\varepsilon_1\hat{k}_t\ssum_{a=1}^3\hat{k}_a^2
+\frac{8\varepsilon_1}{\hat k_t}\ssum_{1\leq a<b \leq3}\hat{k}_a^2\hat{k}_b^2
\right]+\mathcal{O}\big(\tau,\varepsilon^0\big),
\label{eq:bispecctrum3}
\end{align}
which reproduces the result in~\cite{Maldacena:2002vr}.

\subsubsection{Even $d$}
Let us first look at the contribution from the terms with $\cI_{[-\tau,\infty)}$ in \eqref{eq:Itransformed} and \eqref{eq:Idottransformed}. 
Since both these integrals and $i^d$ are real, we need the real part of $\cH^{(1)}$. Since the real part of $H_\nu^{(1)}(z)$ $(z>0)$ starts at order $z^\nu$ for small $z$ (see the property (1) at the beginning of this subsection), the real part of $\cH^{(1)}$ at late time is of order
\begin{align}
\text{Re}[\cH^{(1)}(\tau)] = \cO((-\tau)^d,(-\tau)^d\ln(-\tau))\,.
\end{align}
On the other hand, the leading terms of the two integrals $\cI_{[-\tau,\infty)}^{(\frac{d}{2}+1,\frac{d}{2},\frac{d}{2},\frac{d}{2})},\cI_{[-\tau,\infty)}^{(\frac{d}{2}+1,\frac{d}{2},\frac{d}{2}-1,\frac{d}{2}-1)}$ are of order $(-\tau)^{2-d},(-\tau)^{4-d}$, respectively, according to \eqref{cI1-leading} and \eqref{cI2-leading}. Therefore, \emph{the $\cI_{[-\tau,\infty)}$ parts in \eqref{eq:Itransformed} and \eqref{eq:Idottransformed} vanish in the late time limit}.

Let us proceed to the parts with $\cI_{\ga_2}$. The computation is parallel to the odd $d$ case. For concreteness, we consider the case $d=4$, for which $\cH^{(1)}$ is expanded as 
\begin{align}
\begin{split}
&\cH^{(1)}(\tau)|_{d=4}
=\frac{i}{\pi^3 \hat{k}_1^2 \hat{k}_2^2 \hat{k}_3^2}\bigg(64+16\big(\hat{k}_1^2+\hat{k}_2^2+\hat{k}_3^2\big)\tau^2-\Big[(3-2i\pi)\big(\hat{k}_1^4+\hat{k}_2^4+\hat{k}_3^4\big) \\
&\qquad\quad -2\big(\hat{k}_1^2+\hat{k}_2^2+\hat{k}_3^2\big)^2 +4\big(\gamma_E-1+\ln(-\tau\hat{k}_1/2)\big)\hat{k}_1^4 \\
&\qquad\quad +4\big(\gamma_E-1+\ln(-\tau\hat{k}_2/2)\big)\hat{k}_2^4+4\big(\gamma_E-1+\ln(-\tau\hat{k}_3/2)\big)\hat{k}_3^4\Big]\tau^4\bigg)+\mathcal{O}\big(\tau^6\big).
\end{split}
\end{align}
We then obtain
\begin{align}
\text{Re}\left[i^{6}\cH^{(1)}(\tau)\cI_{\gamma_2}^{(3,2,2,2)}\right]
&=\frac{64\big(\hat{k}_1^2+\hat{k}_2^2+\hat{k}_3^2\big)}{\pi^2\hat{k}_1^4 \hat{k}_2^4 \hat{k}_3^4} + \mathcal{O}(\tau), \label{ga2_I1d4} \\
\text{Re}\left[i^{4}\cH^{(1)}(\tau)\cI_{\gamma_2}^{(3,2,1,1)}\right] 
&=\frac{64}{\pi^2\hat{k}_1^4\hat{k}_2^2\hat{k}_{3}^2}+\mathcal{O}(\tau). \label{ga2_I2d4}
\end{align}
Summing up all contributions, we obtain
\begin{align}
I_{\hat{k}_1\hat{k}_2\hat{k}_3}
&= -\frac{16iH^{9}\big(\hat{k}_1^2+\hat{k}_2^2+\hat{k}_3^2\big)}{27\pi^2\varepsilon_1^3\hat{k}_1^4 \hat{k}_2^4 \hat{k}_3^4}+\mathcal{O}(\tau)\,,\\
I^\bullet_{\hat{k}_1\hat{k}_2\hat{k}_3}
&= -\frac{16iH^{11}}{27\pi^2\hat{k}_1^4\hat{k}_2^{2}\hat{k}_3^2\varepsilon_1^3}+\mathcal{O}(\tau)\,.
\end{align}
Plugging them back into \eqref{I1def}--\eqref{I3def}, we obtain $I_1(\hat t),I_2(\hat t),I_3(\hat t)$:
\begin{align}
\begin{split}
I_1(\hat{t})|_{d=4}
&=\frac{16H^6}{9\pi^2\varepsilon_1\hat{k}_1^4\hat{k}_2^4\hat{k}_3^4}\ssum_{1\leq a<b\leq 3}\hat{k}_a^2\hat{k}_b^2+\mathcal{O}\big(\tau,\varepsilon^0\big),
\end{split}\\
\begin{split}
I_2(\hat{t})|_{d=4}
&=\frac{8H^6}{9\pi^2\varepsilon_1\hat{k}_1^4\hat{k}_2^4\hat{k}_3^4}\left(\ssum_{a=1}^3\hat{k}_a^2\right)^2+\mathcal{O}\big(\tau,\varepsilon^0\big),
\end{split}\\
\begin{split}
I_3(\hat{t})|_{d=4}
&=\frac{16H^6}{9\pi^2\varepsilon_1\hat{k}_1^4\hat{k}_2^4\hat{k}_3^4}\left[
4\ssum_{1\leq a<b\leq 3}\hat{k}_a^2\hat{k}_b^2-\left(\ssum_{a=1}^3\hat{k}_a^2\right)^2\right]+\mathcal{O}\big(\tau,\varepsilon^0\big).
\end{split}
\end{align}
Finally, summing up $I_0,I_1,I_2,I_3$, we obtain
\begin{align}
\mathcal{B}_{\hat k_1,\hat k_2,\hat k_3}^{d=4}
&=\frac{8H^6}{9\pi^2\varepsilon_1^2\hat{k}_1^4\hat{k}_2^4\hat{k}_3^4}
\left[(\varepsilon_2-\varepsilon_1)\ssum_{a=1}^3\hat{k}_a^4+8\varepsilon_1\ssum_{1\leq a<b\leq 3}\hat{k}_a^2\hat{k}_b^2\right]
+\mathcal{O}\big(\tau,\varepsilon^0\big).
\label{eq:bispectrum4}
\end{align}

Its squeezed limit $\hat k_1\ll\hat k_2\sim\hat k_3$ reads
\begin{align}
    \mathcal{B}_{\hat k_1,\hat k_2,\hat k_3}^{d=4}\Big|_{\hat{k}_1\ll\hat k_2}=\frac{16H^6}{9\pi^2\varepsilon_1^2\hat{k}_1^4\hat{k}_2^4}(3\varepsilon_1+\varepsilon_2)+\mathcal{O}\left(\tau,\varepsilon^0,\frac{\hat{k}_1}{\hat{k}_2}\right).
\end{align} 
One can verify by using the power spectrum \eqref{eq:Platetime} with $d=4$ that it satisfies the consistency relation generalised to $d=4$ given by:
\begin{align}
\label{eq:consistency1}
\mathcal{B}_{\hat k_1,\hat k_2,\hat k_3}^{d=4}\Big|_{\hat{k}_1\ll\hat k_2}=-\mathcal{P}^{d=4}_{\hat{k}_1}\left(4+\hat{k}_2\frac{\rd}{\rd\hat{k}_2}\right)\mathcal{P}^{d=4}_{\hat{k}_2}+\mathcal{O}\left(\frac{\hat{k}_1}{\hat{k}_2}\right).
\end{align}

\subsection{Bispectrum on the brane}

Let us consider the three-point function localised on the 3-brane:
\begin{align}
\bra\zt(\hat{t},\bsx_1,\bsnl)\zt(\hat{t},\bsx_2,\bsnl)\zt(\hat{t},\bsx_3,\bsnl)\ket \,.
\end{align}
Similarly to the power spectrum, it can be rewritten as a Fourier transform on the brane and the sum over KK modes:\footnote{Here we introduced the shorthand as for the general dimensional one:
\begin{align}
\int_{\bsk_1,\bsk_2,\bsk_3}:=\int\frac{\rd^3\bsk_1}{(2\pi)^3}\frac{\rd^3\bsk_2}{(2\pi)^3}\frac{\rd^3\bsk_3}{(2\pi)^3},
\end{align}}
\begin{align}
\begin{split}
\bra\zt(\hat{t},\bsx_1,\bsnl)\zt(\hat{t},\bsx_2,\bsnl)\zt(\hat{t},\bsx_3,\bsnl)\ket
=\int_{\bsk_1,\bsk_2,\bsk_3}e^{i\sum_{a=1}^3\bsk_a\cdot\bsx_a}(2\pi)^3\de^3\left(\ssum_{a=1}^3\bsk_a\right)
\cB_{k_1,k_2,k_3}^{\mrm{brane}},
\end{split}
\end{align}
where the bispectrum on the brane $\cB_{k_1,k_2,k_3}^{\mrm{brane}}$ is given by
\begin{align}
\cB_{k_1,k_2,k_3}^{\mrm{brane}}
=\frac{1}{(2\pi R_0)^{2d-6}}\sum_{\bsn_1+\bsn_2+\bsn_3=0} 
\mathcal{B}_{\hat{k}_1(\bsn_1),\hat{k}_2(\bsn_2),\hat{k}_3(\bsn_3)},
\end{align}
where the sum is taken over $\bsn_1,\bsn_2,\bsn_3\in\bbZ^{d-3}$ under the constraint $\bsn_1+\bsn_2+\bsn_3=0$.

Let us consider the case $d=4$. The bispectrum on the brane can then be rewritten as
\begin{align}
\begin{split}
\cB_{k_1,k_2,k_3}^{\mrm{brane}}
&=\frac{1}{(2\pi R_0)^2}\frac{8H^6}{9\pi^2\varepsilon_1^2}
\bigg\{(\varepsilon_2-\varepsilon_1)\Big(S'_{k_1,k_2,k_3}+S'_{k_2,k_3,k_1}+S'_{k_3,k_1,k_2}\Big) \\
&\qquad\qquad\qquad\qquad\quad
+8\varepsilon_1\Big(S''_{k_1,k_2,k_3}+S''_{k_2,k_3,k_1}+S''_{k_3,k_1,k_2}\Big)\bigg\}+\mathcal{O}\big(\tau,\varepsilon^0\big)\,,
\end{split}\label{Bbrane-general}
\end{align}
where we introduced two series $S',S''$ as
\begin{align}
    S'_{k_1,k_2,k_3}&:=\sum_{n_1+n_2+n_3=0}\frac{1}{\hat{k}_2(n_2)^4\hat{k}_3(n_3)^4},\\
    S''_{k_1,k_2,k_3}&:=\sum_{n_1+n_2+n_3=0}\frac{1}{\hat{k}_1(n_1)^4\hat{k}_2(n_2)^2\hat{k}_3(n_3)^2}\,.
\end{align}
These series have analytic expressions:
\begin{align}
\begin{split}
S'_{k_1,k_2,k_3}
&=\frac{\pi^2 R_0^2}{16k_2^3k_3^3}\text{csch}(\pi R_0k_2)^2\text{csch}(\pi R_0k_3)^2\\
&\qquad\times\big(\sinh(2\pi R_0k_2)+2\pi R_0k_2\big)\big(\sinh(2\pi R_0k_3)+2\pi R_0k_3\big)
\end{split}
\end{align}
and
\begin{align}
\begin{split}
    &S''_{k_1,k_2,k_3}=\pi^2R_0^2\bigg[A_{k_1,k_2,k_3}+B_{k_1,k_2,k_3}\coth(\pi R_0k_2)\coth(\pi R_0k_3)\\
    &\qquad+\coth(\pi R_0k_2)k_3\Big(C_{k_1,k_2,k_3}\coth(\pi R_0k_1)+\pi R_0k_1D_{k_1,k_2,k_3}\text{csch}(\pi R_0k_1)^2\Big)\\
    &\qquad+\coth(\pi R_0k_3)k_2\Big(C_{k_1,k_3,k_2}\coth(\pi R_0k_1)+\pi R_0k_1D_{k_1,k_3,k_2}\text{csch}(\pi R_0k_1)^2\Big)\bigg],
\end{split}
\end{align}
where we have defined
\begin{align}
    A_{k_1,k_2,k_3}&=4\frac{k_1^2-k_2^2-k_3^2}{E_{k_1,k_2,k_3}^2},\\
    B_{k_1,k_2,k_3}&=2k_1^3\frac{E_{k_1,k_2,k_3}+8k_2^2 k_3^2}{F_{k_1,k_2,k_3}},\\
    C_{k_1,k_2,k_3}&=\frac{G_{k_1,k_2,k_3}\big(3k_1^2+k_2^2-k_3^2\big)-8k_1^2k_2^2\big(k_1^2-k_2^2+k_3^2)}{F_{k_1,k_2,k_3}},\\
    D_{k_1,k_2,k_3}&=\frac{G_{k_1,k_2,k_3}\big(k_1^2+k_2^2-k_3^2\big)}{F_{k_1,k_2,k_3}},\\
    E_{k_1,k_2,k_3}&=k_1^4+k_2^4+k_3^4-2\big(k_1^2k_2^2+k_1^2k_3^2+k_2^2k_3^2\big),\\
    F_{k_1,k_2,k_3}&=2k_1^3k_2k_3E_{k_1,k_2,k_3}^2,\\
    G_{k_1,k_2,k_3}&=(-k_1+k_2+k_3)(k_1-k_2+k_3)(k_1+k_2-k_3)(k_1+k_2+k_3).
\end{align}

Let us look at the following three limits:
\begin{align}
&(R_0k_1\ll1, \quad R_0k_2\ll1, \quad R_0k_3\ll1)\,, \\
&(R_0k_1\ll1, \quad R_0k_2\gg1, \quad R_0k_3\gg1)\,, \\
&(R_0k_1\gg1, \quad R_0k_2\gg1, \quad R_0k_3\gg1) \,.
\end{align}
In these limits, the bispectrum on the brane behaves as ($k_t:=k_1+k_2+k_3$),
\begin{align}
&(2\pi R_0)^2\cB_{k_1,k_2,k_3}^{\mrm{brane}}\nonumber\\
&\quad\underset{\substack{R_0k_1\ll1\\R_0k_2\ll1\\R_0k_3\ll1}}{\simeq}
\frac{8H^6}{9\pi^2\varepsilon_1^2k_1^4k_2^4k_3^4}
\Big[
(\varepsilon_2-\varepsilon_1)\ssum_{a=1}^3k_a^4
+8\varepsilon_1\ssum_{1\leq a<b\leq 3}k_a^2k_b^2
\Big],\label{eq:bispectruma}\\
\begin{split}
&\quad\underset{\substack{R_0k_1\ll1\\R_0k_2\gg1\\R_0k_3\gg1}}{\simeq}
\frac{4R_0H^6}{9\pi\varepsilon_1^2k_1^4k_2^3k_3^3}
\Bigg[(\varepsilon_2-\varepsilon_1)\big(k_2^3+k_3^3\big)+16\varepsilon_1\frac{k_2^2k_3^2}{k_2+k_3}\Bigg],\label{eq:bispectrumb}
\end{split}\\
\begin{split}
&\quad\underset{\substack{R_0k_1\gg1\\R_0k_2\gg1\\R_0k_3\gg1}}{\simeq}
\frac{2R_0^2H^6}{9\varepsilon_1^2k_1^3k_2^3k_3^3}
\Bigg[(\varepsilon_2-\varepsilon_1)\ssum_{a=1}^3k_a^3 
+16\varepsilon_1\ssum_{1\leq a<b\leq 3}\left(\frac{k_a^2k_b^2}{k_t}
+k_1k_2k_3\frac{k_ak_b}{k_t^2}\right)\Bigg].
\label{eq:bispectrumc}
\end{split}
\end{align}

In the third limit, the bispectrum differs from the $d=3$ bispectrum \eqref{eq:bispecctrum3} while it is scale invariant. This is in contrast with the power spectrum for which the corresponding limit recovers the $d=3$ power spectrum. It can be understood in terms of three-dimensional global conformal transformations. At the leading order of the Hubble flow parameters, both power spectrum and bispectrum are dilatation invariant. This fixes the momentum dependence of the power spectrum while this is not enough to fix the (shape of the) bispectrum. This is because invariance under special conformal transformations is broken by the inflationary background. Therefore, the bispectra in the two cases above do not have to coincide. This is shown explicitly in Appendix B.

Let us further look at the 3-brane squeezed limit $k_1 \ll k_2\sim k_3$ in each of the three cases above:
\begin{align}
(2\pi R_0)^2\cB_{k_1,k_2,k_3}^{\mrm{brane}}
&\underset{R_0k_1 \ll R_0k_2 \ll 1}{\simeq}
\frac{16H^6}{9\pi^2\varepsilon_1^2k_1^4k_2^4}(3\varepsilon_1+\varepsilon_2),
\label{eq:squeeze3a}\\
(2\pi R_0)^2\cB_{k_1,k_2,k_3}^{\mrm{brane}}
&\underset{R_0k_1 \ll 1 \ll R_0k_2}{\simeq}
\frac{8R_0H^6}{9\pi\varepsilon_1^2k_1^4k_2^3}(3\varepsilon_1+\varepsilon_2),
\label{eq:squeeze3b}\\
(2\pi R_0)^2\cB_{k_1,k_2,k_3}^{\mrm{brane}}
&\underset{1 \ll R_0k_1 \ll R_0k_2}{\simeq}
\frac{4R_0^2H^6}{9\varepsilon_1^2k_1^3k_2^3}(3\varepsilon_1+\varepsilon_2)\,.
\label{eq:squeeze3c}
\end{align}
They can be understood from a consistency relation on the 3-brane for $k_1\ll k_2\sim k_3$, which is obtained by taking the sum of the 5D consistency relation \eqref{eq:consistency1} over $n_1,n_2,n_3$ under the constraint $n_1+n_2+n_3=0$ that reads\footnote{Since the bulk consistency relation is valid for $\hat k_1\ll\hat k_2\sim\hat k_3$, modes such that $|n_1|\gg|n_2|$ whereby $\hat k_1(n_1)\sim\hat k_2(n_2)$ would invalidate the use of the consistency relation in the KK sum. However, these modes are negligible since such $n_1$'s are too large to contribute to the sum of $\cP^{d=4}_{\hat k_1(n_1)}$ over $n_1$ on the right hand side of \eqref{eq:consistency1}.}
\begin{align}
\label{cr-brane}
\cB_{k_1,k_2,k_3}^{\mrm{brane}}\Big|_{k_1\ll k_2\simeq k_3}\simeq(3\vep_1+\vep_2)\cP_{k_1}^{\mrm{brane}}\cP_{k_2}^{\mrm{brane}}.
\end{align}
In the bulk consistency relation \eqref{eq:consistency1}, the momenta satisfy $\hat k_1\ll\hat k_2\sim\hat k_3$ but $\hat k_1$ is not zero. Therefore, the 3D momentum $k_1$ can be much smaller than $\hat k_1$, and for such $k_1$ there are many KK modes between $k_1$ and $\hat k_1$ and hence we still have to take the KK sum over $n_1$.
Once \eqref{cr-brane} is given, its right hand side can be evaluated in each limit given in \eqref{eq:squeeze3a}--\eqref{eq:squeeze3c} by using the asymptotic formulas of $S_\nu$ \eqref{Salpha_limits}, and one can show that \eqref{cr-brane} reproduces the brane squeezed limits \eqref{eq:squeeze3a}--\eqref{eq:squeeze3c}. Recall that when $R_0k_1\ll1$, the sum of $\cP_{\hat k_1(n_1)}$ over $n_1$ is equivalent to just picking up the zero mode, as commented below \eqref{Salpha_limits}.
The common factor $3\vep_1+\vep_2$ is $1-n_s$ \eqref{eq:spectralindexsmall} with the spectral index $n_s$ at the leading order of the Hubble flow parameters, which originates from the \emph{bulk} scalar spectral index of the 5D power spectrum. Indeed, this factor in \eqref{cr-brane} comes from the power $-2\nu\simeq-4-3\vep_1-\vep_2$ in the 5D ($d=4$) power spectrum \eqref{eq:Platetime} plus the factor 4 in the 5D consistency relation \eqref{eq:consistency1}.

Note that in the second limit \eqref{eq:squeeze3b}, the bispectrum gets enhanced compared to the 4D inflation case due to the change of the power of $k_1$ in the denominator ($k_1^4$ from $k_1^3$), similarly to the change of behaviour in the power spectrum at large distances that changes the spectral index from 1 to 0. This enhancement, if this region is observable without large errors, can easily represent 2-3 orders of magnitude, possibly within the reach of future experiments.

\section{Non-gaussianities and concluding remarks}
In this section, we use our results to make predictions for the CMB observations.
From Section~3.3, the power spectrum of 5D inflation $(d=4)$ to leading order in the Hubble flow parameters behaves as
\begin{align}
    (2\pi R_0)\mathcal{P}_{k,\text{brane}}\underset{R_0k\ll1}{\simeq}&\frac{4H^3}{3\pi\varepsilon_1k^4}
    \simeq \frac{2}{R_0k}\mathcal{P}_{k}^{d=3}\,,\\
    (2\pi R_0)\mathcal{P}_{k,\text{brane}}\underset{R_0k\gg1}{\simeq}&\frac{2H^3R_0}{3\varepsilon_1 k^3}
    \to (2\pi R_0)\mathcal{P}_{k}^{d=3}\,, 
\label{PS3}
\end{align}
where \eqref{PS3} is normalised to the observed value given in 4D inflation in \eqref{eq:Platetime} for $d=3$ and \eqref{Pk}~\cite{Anchordoqui:2023etp,Anchordoqui:2024amx}.
In a similar way, we may define the dimensionless momentum dependent non-gaussianity parameter $f_{\text{NL}}$ as~\cite{Maldacena:2002vr}:
\begin{align}
\label{fNLdef1}
    f_{\text{NL}}\equiv\frac{5}{6}\frac{\mathcal{B}_{\text{brane}}(k_1,k_2,k_3)}{\mathcal{P}_{k_1}^{d=3}\mathcal{P}_{k_2}^{d=3}
    +\mathcal{P}_{k_1}^{d=3}\mathcal{P}_{k_3}^{d=3}+\mathcal{P}_{k_2}^{d=3}\mathcal{P}_{k_3}^{d=3}}\,.
\end{align}
Therefore, we find:
\begin{align}
    f_{\text{NL}}&\underset{\substack{R_0k_1\ll1\\R_0k_2\ll1\\R_0k_3\ll1}}{\simeq}\frac{5}{12}\frac{8}{R_0^3k_1k_2k_3}
    \left\{(\varepsilon_2-\varepsilon_1)\frac{k_1^4+k_2^4+k_3^4}{k_1^3+k_2^3+k_3^3}
    +8\varepsilon_1\frac{k_1^2k_2^2+k_1^2k_3^2+k_2^2k_3^2}{k_1^3+k_2^3+k_3^3}\right\}\,,\\
    f_{\text{NL}}&\underset{\substack{R_0k_1\ll1\\R_0k_2\gg1\\R_0k_3\gg1}}{\simeq}\frac{5}{12}\frac{2}{R_0k_1}
    \left\{(\varepsilon_2-\varepsilon_1)+16\varepsilon_1\frac{k_2^2k_3^2}{(k_2+k_3)\big(k_2^3+k_3^3\big)}\right\}\,,\\    
    f_{\text{NL}}&\underset{\substack{R_0k_1\gg1\\R_0k_2\gg1\\R_0k_3\gg1}}{\simeq}\frac{5}{12}\Bigg\{
    (\varepsilon_2-\varepsilon_1)+\frac{16\varepsilon_1}{k_1^3+k_2^3+k_3^3}
    \ssum_{1\leq a<b\leq 3}\Bigg(\frac{k_a^2k_b^2}{k_t}
    +k_1k_2k_3\frac{k_ak_b}{k_t^2}
    \Bigg)\Bigg\}\,,
\label{fNL4Dlimit}
\end{align}
where $k_t=k_1+k_2+k_3$.

Obviously $f_{\text{NL}}$ is not a unique way to normalise the bispectrum. We may also use as normalisation of the observed (brane) bispectrum the power spectrum on the brane as
\begin{align}
    \hat f_{\text{NL}}\equiv\frac{5}{6}\frac{\mathcal{B}_{\text{brane}}(k_1,k_2,k_3)}{\mathcal{P}_{k_1,\text{brane}}\mathcal{P}_{k_2,\text{brane}}
    +\mathcal{P}_{k_1,\text{brane}}\mathcal{P}_{k_3,\text{brane}}+\mathcal{P}_{k_2,\text{brane}}\mathcal{P}_{k_3,\text{brane}}}\,.            
\end{align}
Its limits then turn out to be independent of the compactification radius:
\begin{align}
    \hat f_{\text{NL}}&\underset{\substack{R_0k_1\ll1\\R_0k_2\ll1\\R_0k_3\ll1}}{\simeq}\frac{5}{12}
    \left\{(\varepsilon_2-\varepsilon_1)
    +8\varepsilon_1\frac{k_1^2k_2^2+k_1^2k_3^2+k_2^2k_3^2}{k_1^4+k_2^4+k_3^4}\right\}\,,\\
    \hat f_{\text{NL}}&\underset{\substack{R_0k_1\ll1\\R_0k_2\gg1\\R_0k_3\gg1}}{\simeq}\frac{5}{12}
    \left\{(\varepsilon_2-\varepsilon_1)+16\varepsilon_1\frac{k_2^2k_3^2}{(k_2+k_3)\big(k_2^3+k_3^3\big)}\right\}\,,\\    
    \hat f_{\text{NL}}&\underset{\substack{R_0k_1\gg1\\R_0k_2\gg1\\R_0k_3\gg1}}{\simeq}\frac{5}{12}\Bigg\{
    (\varepsilon_2-\varepsilon_1)+\frac{16\varepsilon_1}{k_1^3+k_2^3+k_3^3}
    \ssum_{1\leq a<b\leq 3}\Bigg(\frac{k_a^2k_b^2}{k_t}
    +k_1k_2k_3\frac{k_ak_b}{k_t^2}
    \Bigg)\Bigg\}\,.
\end{align}
Note that the third limit, namely the limit of small wave lengths compared to the compactification radius, coincides with the same limit \eqref{fNL4Dlimit} of the other normalisation of $f_{\text{NL}}$ \eqref{fNLdef1}.
It may also be compared with the dimensionless non-gaussianity of the 4D single field inflation~\cite{Maldacena:2002vr}:
\begin{align}
    f_{\text{NL}}^{\text{4D}}
    &=\frac{5}{6}\frac{\mathcal{B}^{d=3}(k_1,k_2,k_3)}{\mathcal{P}_{k_1}^{d=3}\mathcal{P}_{k_2}^{d=3}
    +\mathcal{P}_{k_1}^{d=3}\mathcal{P}_{k_3}^{d=3}+\mathcal{P}_{k_2}^{d=3}\mathcal{P}_{k_3}^{d=3}} \nn\\
    &=\frac{5}{12}\Bigg\{
    (\varepsilon_2-\varepsilon_1)+\frac{\varepsilon_1}{k_1^3+k_2^3+k_3^3}
    \ssum_{1\leq a<b\leq 3}\Bigg(8\frac{k_a^2k_b^2}{k_t}+k_ak_b^2+k_a^2k_b
    \Bigg)\Bigg\},
\end{align}
where we used \eqref{eq:Platetime} with $\nu_{d=3}\simeq3/2$ and \eqref{eq:bispecctrum3} (note that $\hat\bsk_a=\bsk_a$ for $d=3$). As mentioned above, $f_{\text{NL}}^{\text{4D}}$ differs from \eqref{fNL4Dlimit} due to the breaking of global conformal invariance.

Since the expressions of physical observables to the Hubble flow parameters depend on the number of spatial dimensions [see~\eqref{eq:spectralindexsmall}], we give here the relations of $\varepsilon_1$ and $\varepsilon_2$ of 5D inflation $(d=4)$ to the tilt of the scalar power spectrum $(n_s^\text{obs}-1)$ and the tensor-to-scalar ratio $r^\text{obs}$~\cite{Antoniadis:2023sya}:
\begin{align}
n_s^\text{obs}-1= -(3\varepsilon_1+\varepsilon_2)\,;~~ r^\text{obs}=24\varepsilon_1 \quad\Rightarrow\quad
8\varepsilon_1=\frac{r^{\text{obs}}}{3}\,;~~ \varepsilon_2-\varepsilon_1=\frac{r^{\text{obs}}}{12}+1-n_s^\text{obs}\,.
\end{align}

In summary, in this paper we have computed the bispectrum observed on a 3-brane, to leading order in slow-roll single field inflation in five dimensions. Similarly to the power spectrum, scale invariance is restored for large 3D momenta corresponding to angular distances less than about 10 degrees in the sky but the shape dependence is different from 4D inflation due to the breaking of invariance under global special conformal transformations. Moreover, there is an important enhancement in the squeezed limit that could be a smoking gun of higher dimensional inflation, if it can be measured experimentally without large errors.

\section*{Acknowledgements}

I.A. and H.I. are supported by the Second Century Fund (C2F), Chulalongkorn University. This research is funded by Thailand Science Research and Innovation Fund Chulalongkorn University, grant number IND\_FF\_68\_369\_2300\_097. We are grateful to Spyros Sypsas, Shinji Tsujikawa and Vicharit Yingcharoenrat for valuable discussions.

\appendix

\section{Fluctuations and their cubic action in ADM formalism}

This appendix presents basic formulas in the ADM formalism in general dimensions and the cubic action of the scalar fluctuation in the comoving gauge in a single field slow-roll inflation model.

\subsubsection*{Brief overview of ADM formalism}

Let us suppose that the spacetime has $(d+1)$-dimensions. In ADM formalism, the metric is parametrised as ($t=x^0$)\footnote{In this appendix we use $x^\mu=(t,x^i)$, which differs from the notation in the main text.}
\begin{align}
\rd s^2=g_{\mu\nu}\rd x^\mu\rd x^\nu
=-N^2\rd t^2+a(t)^2h_{ij}(\rd x^i+N^i\rd t)(\rd x^j+N^j\rd t),
\end{align}
where $N$ is the lapse and $N^i$ is the shift vector. 
Latin indices, which run over spatial directions, are raised/lowered with the spatial metric $h_{ij}$ and its inverse $h^{ij}$. The scalar Riemann cuvature $R(g)$ is expressed in terms of $N,N_i,h_{ij}$ as
\begin{align}
&R(g)=R(h)+E_{ij}E^{ij}-E^2, \\
&E_{ij}=\tfrac{1}{2}N(\dot h_{ij}-\ol{\nab}_iN_j-\ol{\nab}_jN_i), \qquad
E=h^{ij}E_{ij}, \qquad E^{ij}=h^{ik}h^{j\ell}E_{k\ell}\,,
\end{align}
where $R(h)$ is the scalar Riemann curvature of $h$, and $\ol\nab$ is the covariant derivative with the Christoffel connection of $h$. The dot means $\pd/\pd t$.

A typical two-derivative action of a scalar field minimally coupled with the Einstein-Hilbert action reads (using $-g=N^2h$)
\begin{align}
\begin{split}
    S&=\int \rd t\rd^dx\sqrt{-g}\left\{\frac{1}{2}R-\frac{1}{2}g^{\mu\nu}\partial_\mu\phi\partial_\nu\phi-V(\phi)\right\}\\
    &=\frac{1}{2}\int \rd t\rd^dx\sqrt{h}N\left\{\hat{R}+\frac{1}{N^2}(E_{ij}E^{ij}-E^2)+\frac{1}{N^2}(\dot{\phi}-N^i\partial_i\phi)^2-h^{ij}\partial_i\phi\partial_j\phi-2V\right\}\,.
    \end{split}
\label{S-ADM}
\end{align}
A striking feature is that the lapse and shift vectors $N$ and $N^i$ satisfy the algebraic equations of motion ($E_i^j=E_{ik}h^{kj}$):
\begin{align}
    0&=R(h)-N^{-2}(E_{ij}E^{ij}-E^{2})-N^{-2}(\dot{\phi}-N^i\partial_i\phi)^2-h^{ij}\partial_i\phi\partial_j\phi-2V,\label{eq:constraint1}\\
    0&=\ol{\nabla}_j\left[N^{-1}(E_i^j-\delta_i^jE)\right]-N^{-1}(\dot{\phi}-N^j\partial_j\phi)\partial_i\phi.\label{eq:constraint2}
\end{align}
One then solves these equations for $N,N_i$ order by order in fluctuations in some gauge. Substituting the solutions into the action \eqref{S-ADM} gives the action for the fluctuations.

\subsubsection*{Cubic action at full order in Hubble flow parameters}

We give two expressions of the interaction Lagrangians cubic in $\zt$ at full order in Hubble flow functions. The first corresponds to the one given in~\cite{Maldacena:2002vr} for $d=3$:
\begin{align}
M_*^{1-d}\cL_3&=
\frac{d-1}{2}a^d\vep_1^2\zt\dot\zt^2
+\frac{d-1}{2}a^{d-2}\vep_1^2\zt(\pd_i\zt)^2
-(d-1)a^d\vep_1^2\dot\zt\pd_i\zt\pd^{-2}\pd_i\dot\zt \nn\\
&\quad
+(d-1)[\pd_t(a^d\vep_1\dot\zt)-a^{d-2}\vep_1\pd^2\zt]f(\zt) \nn\\
&\quad
+\frac{d-1}{4}a^dH\vep_1\vep_2\vep_3\zt^2\dot\zt
+\frac{1}{4}a^d\vep_1^3(\zt\pd^{-2}\pd_i\pd_j\dot\zt\pd^{-2}\pd_i\pd_j\dot\zt-\zt\dot\zt^2),
\label{fullaction1}
\end{align}
while the second corresponds to the one given in~\cite{Burrage:2011hd} for $d=3$:
\begin{align}
M_*^{1-d}\cL_3&=
\frac{d-1}{2}a^d(\vep_1^2-\vep_1\vep_2)\zt\dot\zt^2
+\frac{d-1}{2}a^{d-2}(\vep_1^2+\vep_1\vep_2)\zt(\pd_i\zt)^2
-(d-1)a^d\vep_1^2\dot\zt\pd_i\zt\pd^{-2}\pd_i\dot\zt \nn\\
&\quad
+(d-1)[\pd_t(a^d\vep_1\dot\zt)-a^{d-2}\vep_1\pd^2\zt]\left[f(\zt)-\frac{\vep_2}{4}\zt^2\right] \nn\\
&\quad
+\frac{1}{2}a^d\vep_1^3\dot\zt\pd_i\zt\pd^{-2}\pd_i\dot\zt
+\frac{1}{4}a^d\vep_1^3\pd^2\zt\pd^{-2}\pd_i\dot\zt\pd^{-2}\pd_i\dot\zt\,.
\label{fullaction2}
\end{align}
These two actions differ by temporal total derivative terms. The difference does not contribute to the correlators as long as we adopt the path integral formulation of the in-in formalism.

\section{Bispectrum under conformal transformation}

The isometry of four dimensional de Sitter space is the three-dimensional conformal group. In single field slow-roll inflation models, the background configuration of an inflaton breaks invariance under dilatation and special conformal transformation.
It is therefore natural that the bispectrum of $(1+3)$-dimensional single field inflation is not invariant under three-dimensional special conformal invariance at any order of slow-roll parameters. 
This violation leads to various momentum dependence of bispectra in different inflation models.
In this appendix, we explicitly demonstrate the (approximate) restoration of dilation symmetry and violation of special conformal symmetry, taking as examples the bispectrum \eqref{eq:bispecctrum3} of the simplest setup~\cite{Maldacena:2002vr} and our 3-brane bispectrum \eqref{eq:bispectrumc} obtained by taking KK sum of the 5D one. We also demonstrate the approximate invariance of the power spectra of these setups under dilatation and special conformal transformation up to slow-roll parameters.

Since these power and bispectra are given at late time limit, they are correlators in three-dimensional Euclidean space, which is the future boundary of 4D de Sitter space. Their invariance under dilatation and special conformal transformation in momentum space are dictated by the Ward-Takahashi (WT) identities~\cite{Maldacena:2011nz,Coriano:2013jba,Bzowski:2013sza}. A two-point function is a function of the norm of a single momentum $V(k)$ as a consequence of translation and rotation invariance. If it is a two-point function of scalar fields of conformal dimensions $\De_1,\De_2$, the WT identities for dilatation and special conformal invariance are given by\footnote{In $d$-dimensional case, 3 is replaced by $d$ in \eqref{2pt-dt} and $4-2\De_1$ is replaced by $1+d-2\De_1$ in \eqref{2pt-sct}.}
\begin{align}
&\left(\De_1+\De_2-3-k\frac{\rd}{\rd k}\right)V(k)=0, \label{2pt-dt}\\
&\left[\frac{\rd^2}{\rd k^2}+(4-2\De_1)\frac{1}{k}\frac{\rd}{\rd k}\right]V(k)=0, \label{2pt-sct}
\end{align}
where the first one is for dilatation and the second for special conformal transformation. Solving both gives $V(k)\propto k^{2\De_1-3}$ with $\De_1=\De_2$. 

Since the scalar perturbation $\zt$ has classical conformal dimension 0 and the power spectrum to leading order is $V(k)\propto k^{-3}$ (in $d=3$), both equations are satisfied and the power spectrum is invariant under global conformal transformations.

Let us move on to three-point functions. By translation and rotation invariance, they take the form $V_3(k_1,k_2,k_3)$ where $k_a$ is the norm of momentum $\bsk_a$ and the three momenta satisfy the momentum conservation $\bsk_1+\bsk_2+\bsk_3=\bsnl$. If $V_3$ is a correlator of scalar operators of conformal dimensions $\De_1,\De_2,\De_3$, the WT identities for dilatation and special conformal invariance read\footnote{In $d$-dimensional case, $\De_t-6$ in \eqref{3pt-dil} is replaced by $\De_t-2d$, and $4-2\De_a$ in \eqref{3pt-sct} is replaced by $1+d-2\De_1$.}
\begin{align}
&\left(k_1\frac{\pd}{\pd k_1}+k_2\frac{\pd}{\pd k_2}+k\frac{\pd}{\pd k_3}\right)V_3(k_1,k_2,k_3)=(\De_t-6)V_3(k_1,k_2,k_3), \label{3pt-dil} \\
&\cK_1V_3(k_1,k_2,k_3)=\cK_2V_3(k_1,k_2,k_3)=\cK_3V_3(k_1,k_2,k_3), \label{3pt-sct}
\end{align} 
where $\De_t:=\De_1+\De_2+\De_3$, and $\cK_a$ ($a=1,2,3$) is the second-order differential operator with respect to $k_a$ which has already appeared in \eqref{2pt-sct},
\begin{align}
\cK_a:=\frac{\pd^2}{\pd k_a^2}+(4-2\De_a)\frac{1}{k_a}\frac{\pd}{\pd k_a}. 
\end{align}
Notice that in \eqref{3pt-sct} the common value of $\cK_aV_3$ is not constrained. Their solution is given by
\begin{align}
V_3(k_1,k_2,k_3)=C\int_0^\infty\frac{\rd z}{z}z^{3/2}\sprod_{a=1}^3k_a^{\nu_a}K_{\nu_a}(k_az),
\label{V3-sol}
\end{align}
where $\nu_a:=\De_a-3/2$.
This satisfies \eqref{3pt-sct} because $\cK_a[k_a^{\nu_a}K_{\nu_a}(k_az)]=z^2[k_a^{\nu_a}K_{\nu_a}(k_az)]$. Next, rescaling three $k_a$ is equivalent to rescaling $z$ and the power of $z$ makes \eqref{V3-sol} satisfy \eqref{3pt-dil}.

Let us apply the WT identities to the bispectra \eqref{eq:bispecctrum3} and \eqref{eq:bispectrumc}. First, both satisfy the dilatation WT identity \eqref{3pt-dil} with conformal weights $\De_1=\De_2=\De_3=0$ as expected since they are at leading order in Hubble flow parameters. Next, they do not satisfy the WT identity \eqref{3pt-sct} for the special conformal transformation. In more detail, the terms with the local shape $k_1^3+k_2^3+k_3^3$ in \eqref{eq:bispecctrum3} and \eqref{eq:bispectrumc} satisfy the WT identities\footnote{The equilateral case $k_1=k_2=k_3$ also satisfies the WT identities.} while the other terms do not. Indeed
\begin{align}
\begin{split}
(\cK_1-\cK_2)\cB^{d=3}_{k_1,k_2,k_3}
&=\frac{H^4(k_1-k_2)}{16k_1^4k_2^4k_3^3k_t^2\vep_1}
[k_1^4+2k_1^3(6k_2+k_3)+k_1^2(30k_2+14k_2k_3-6k_3^2) \\
&\quad+2k_1(6k_2^3+7k_2^2k_3-6k_2k_3^2+k_3^3)
+(k_2-k_3)^2(k_2^2+4k_2k_3+k_3^2)],
\end{split} \\ 
\begin{split}
(\cK_1-\cK_2)\cB^{\mathrm{brane}}_{k_1,k_2,k_3}
&=\frac{16H^6(k_1-k_2)}{9\pi^2k_1^3k_2^3k_3^3k_t^3\vep_1}
[k_1^3+k_1^2(4k_2+k_3)+k_1(4k_2+k_3)(k_2+2k_3) \\
&\qquad\qquad\qquad\qquad+k_2(k_2+k_3)(k_2+2k_3)],
\end{split}
\end{align}
where $\cB^{\mathrm{brane}}_{k_1,k_2,k_3}$ stands for \eqref{eq:bispectrumc}.
$(\cK_1-\cK_3)$ on them are obtained by the flip $k_2\leftrightarrow k_3$.

\providecommand{\href}[2]{#2}\begingroup\raggedright\endgroup

\end{document}